\documentclass[12pt,a4paper]{article}
\usepackage{epsfig}
\usepackage{amsfonts}
\usepackage{amssymb}
\usepackage{color}

\begin{document}
%
%
\begin{center}

\vspace{1cm}

{\bf \large ON FORM FACTORS IN $\mathcal{N}=4$ SYM THEORY AND
POLYTOPES} \vspace{2cm}

{\bf \large L. V. Bork$^{1,2}$}\vspace{0.5cm}

{\it $^1$Institute for Theoretical and Experimental Physics, Moscow,
Russia,\\
$^2$The Center for Fundamental and Applied Research, All-Russia
Research Institute of Automatics, Moscow, Russia}\vspace{1cm}

\abstract{In this paper we discuss different recursion relations
(BCFW and all-line shift) for the form factors of the operators from the
$\mathcal{N}=4$ SYM stress-tensor current supermultiplet $T^{AB}$ in
momentum twistor space. We show that cancelations of spurious poles
and the equivalence between different types of recursion relations can
be naturally understood using geometrical interpretation of the form
factors as a special limit of the volumes of polytopes in
$\mathbb{C}\mathbb{P}^4$ in close analogy with the amplitude case. We
also show how different relations for the IR pole coefficients can
be easily derived using the momentum twistor representation. This raises
an intriguing question - which of powerful on-shell methods and ideas
can survive off-shell ?}
\end{center}


\newpage

\tableofcontents{}\vspace{0.5cm}

\renewcommand{\theequation}{\thesection.\arabic{equation}}
\section{Introduction}\label{p1}
In the last years, tremendous progress has been achieved in
understanding the structure of the $S$-matrix (amplitudes) of four dimensional
gauge theories \cite{Reviews_Ampl_General}. The most impressive results have been
obtained in the $\mathcal{N}=4$ SYM theory
(for example, see \cite{Henrietta_Amplitudes} and reference therein).
New computational techniques such
as different sets of recursion relations for the tree level amplitudes and
the unitarity  based methods for loop amplitudes were used to
obtain deep insights in the structure of the $\mathcal{N}=4$ SYM $S$-matrix.
It is believed that these efforts will eventually lead to the complete
determination of the $\mathcal{N}=4$ SYM $S$-matrix in the planar limit. Also,
probably, some beautiful geometrical ideas and insights will be encountered
along the way
\cite{Hoges_Polytopes,Masson_Skiner_Grassmaians_Twistors,Arcani_Hamed_Polytopes,Arcani_Hamed_PositiveGrassmannians,Arcani_Hamed_Amplituhdron_1,Arcani_Hamed_Amplituhdron_2}.

There is another class of objects of interest in $\mathcal{N}=4$ SYM which
resembles amplitudes - the form factors. The form factors are
the matrix elements of the form
\begin{equation}
\langle p_1^{\lambda_1}, \ldots,
p_n^{\lambda_n}|\mathcal{O}|0\rangle,
\end{equation}
where $\mathcal{O}$ is some gauge invariant operator which acts on
the vacuum of the theory and produces some state $\langle p_1^{\lambda_1}, \ldots,
p_n^{\lambda_n}|$ with momenta $p_1, \ldots, p_n$ and
helicities $\lambda_1, \ldots, \lambda_n$\footnote{Note that
scattering amplitudes in ''all ingoing" notation can schematically
be written as $\langle p_1^{\lambda_1}, \ldots, p_n^{\lambda_n}|0
\rangle$. }. One
can think about this object as an amplitude of the proses where
classical current or field, coupled via a gauge invariant
operator $\mathcal{O}$, produces some
quantum state $\langle p_1^{\lambda_1}, \ldots,p_n^{\lambda_n}|$.

It is interesting to study the form factors in $\mathcal{N}=4$ SYM systematically
for several reasons:
\begin{itemize}
\item Symmetries, such as dual conformal
symmetry, play an essential role in the structure of amplitudes in gauge theories.
Moreover, it is expected that $\mathcal{N}=4$ SYM
is an integrable system (see 
\cite{BeisertYangianRev,Staudacher_SpectralReg_New,Beisert_SpectralReg_New,Derkachev_SpectralReg_New} 
and references there, also see \cite{deleeuwm_2014_1,deleeuwm_2014_2}).
Studying the form factors in integrable systems (for example,
see \cite{FF_in_integrable_sys} and references therein) usually
can be useful for a better understanding of the origins
and properties of symmetries in this type of theories.
One may hope that studying the form factors in $\mathcal{N}=4$ SYM
may be useful for understanding
symmetry properties of the $\mathcal{N}=4$ SYM S-matrix
and correlation functions.

\item The form factors are intermediate objects between
fully on-shell quantities such as amplitudes and fully
off-shell quantities such as correlation functions (which are
one of the central objects in AdS/CFT). Since powerful
computational methods have recently appeared for the amplitudes in
$\mathcal{N}=4$ SYM, it would be desirable
to have their analog for the correlation functions
\cite{Raju:2011mp,Roiban_Correlation_Functions}.
Understanding of the structure of the form
factors and the development of computational methods will be
useful for a better understanding of the structure of the correlation functions
of multiple ($n > 2$) gauge invariant local operators in $\mathcal{N}=4$ SYM.
The latter
may also be useful in understanding of "triality" relations: amplitudes,
Willson loops, correlation functions and subsequent relations for
the amplitudes \cite{HuotEquation,Twistors_DescentEquation}.

\item The form factors in $\mathcal{N}=4$ SYM are excellent objects
for developing
and testing new computational methods which can be efficient beyond the
planar sector of maximally supersymmetric gauge theories. Indeed, form factors
naturally incorporate non planarity and violate some supersymmetries
(at least the form factors of the operators from the chiral truncation
of the $\mathcal{N}=4$ SYM stress tensor
supermultiplet).
\end{itemize}

The investigation of the form factors in $\mathcal{N}=4$ SYM was first
initiated in \cite{VanNeerven_1985}, almost 20 years ago. Unique investigation
of form factors of single field non gauge invariant operators
(off-shell currents) was made in \cite{Perturbiner}, by
using the "perturbiner" technique.

After a pause that lasted for nearly a decade the investigation of
1/2-BPS form factors was initiated in
\cite{FormFactorMHV_component_Brandhuber,BKV_Form_Factors_N=4SYM}.
Different on-shell methods
were successfully applied to the form factors
\cite{HarmonyofFF_Brandhuber,BKV_SuperForm,BORK_NMHV_FF,FormFactorMHV_half_BPS_Brandhuber}.
Different multiloop results were obtained in
\cite{BKV_Form_Factors_N=4SYM,FF_MHV_3_2loop,FormFactorMHV_Remainder_half_BPS_Brandhuber}
Different types of regularizations and colour-kinematic duality were considered
in \cite{Henn_Different_Reg_FF,FF_Colour_Kinematic}.
Strong coupling limit results for the form factors were obtained in
\cite{Zhiboedov_Strong_coupling_FF,Strong_coupling_FF_Yang_Gao}.
The form factors in theories with maximal supersymmetry in dimensions
different from $D=4$ were investigated in 
\cite{FF_ABJM_Young,FF_Sudakov_ABJM_Baianchi,Penati_Santambrogio_ABJM_finite_N}.

The aim of this article is the following: we would like to apply the momentum
twistor representation for the form factors of the $\mathcal{N}=4$ SYM stress-tensor
supermultiplet and formulate the BCFW recursion relation
for tree level form factors in this formulation. It is known that in the case
of the amplitudes written in momentum twistor variables, interesting geometrical
properties and symmetries of the amplitudes are represented most clearly
and naturally \cite{Hoges_Polytopes,Arcani_Hamed_Polytopes}.
It is interesting to know what the situation would be if we will consider
partially off-shell object ? What on-shell ideas and methods
such as \cite{Hoges_Polytopes,Arcani_Hamed_Polytopes,Arcani_Hamed_PositiveGrassmannians}
could survive for partially off-shell objects ?

This article is organised as follows. In section 2, we briefly
discuss the general structure of the form factors of the operators from
the $\mathcal{N}=4$ SYM stress-tensor
supermultiplet in on-shell harmonic superspace.
In section 3, we establish and solve BCFW recursion relations for tree
level form factors in the NMHV sector in the on-shell harmonic superspace.
In section 4, we
discuss how to rewrite NMHV form factors in the momentum twistor representation,
establish BCFW recursion relations for general $\mbox{N}^k\mbox{MHV}$ form
factors in the momentum twistor space. In section 5, we represent a sketch of the
proof of the equivalence between BCFW and all-line shift (CSW) recursion relations
for the NMHV sector
in the momentum twistor space and use the geometrical representation of the form
factors as a special limit of the volumes of the polytopes to show that the all-line shift (CSW) representation
of the NMHV sector is free from spurious poles. The latter would imply the spurious
poles cancellation in the BCFW representation as well. In the appendix, we give
more details of the harmonic superspace construction, discuss some particular
examples of the spurious poles cancellation and also discuss how relations
between IR pole coefficients at one loop in the NMHV sector can be naturally
established in the momentum twistor representation.

\section{Form factors of the stress-tensor current
supermultiplet in $\mathcal{N}=4$ SYM}\label{p2}
In this chapter, we are going to introduce essential ideas and notation
regarding the
general structure of the form factor of the stress-tensor supermultiplet formulated
in the harmonic superspace.

To describe the stress-tensor supermultiplet in a manifestly supersymmetric and
$SU(4)_R$ covariant way
it is useful to consider the harmonic superspace parameterized by the set
of coordinates \cite{N=4_Harmonic_SS,SuperCor1}:
\begin{eqnarray}
\mbox{$\mathcal{N}=4$ harmonic
superspace}&=&\{x^{\alpha\dot{\alpha}},
~\theta^{+a}_{\alpha},\theta^{-a'}_{\alpha},
~\bar{\theta}_{a~\dot{\alpha}}^{+},\bar{\theta}_{a'~\dot{\alpha}}^{-},u
\}.
\end{eqnarray}
Here $u$ is the set of
$$
\frac{SU(4)}{SU(2) \times SU(2)' \times U(1)}
$$
harmonic variables, $a$ and $a'$ are the $SU(2)$ indices, $\pm$ corresponds
to $U(1)$ charge; $\theta$'s are Grassmann coordinates,
$\alpha$ and $\dot{\alpha}$ are the $SL(2,\mathbb{C})$ indices.
Hereafter we will not write some indices explicitly in all
expressions when it does not lead to misunderstanding.

The stress-tensor supermultiplet will be given by
\begin{eqnarray}
T=Tr(W^{++}W^{++})
\end{eqnarray}
where $W^{++}(x,\theta^{+},\bar{\theta}^{+})$ is the harmonic
superfield that contains all component fields of the $\mathcal{N}=4$
supermultiplet, which are the $\phi^{AB}$ scalars
(anti-symmetric in the $SU(4)_R$ indices $AB$), $\psi^A_{\alpha},
\bar{\psi}^A_{\dot{\alpha}}$ fermions and $F^{\mu\nu}$ is the gauge
field strength tensor, all in the adjoint representation of
the $SU(N_c)$ gauge group. The details of harmonic superspace construction
will be given in the appendix.
Note that this superfield is on-shell in the sense that algebra
of supersymetric transformations which should
leave $W^{++}$ invariant, is closed only if the component fields in
$W^{++}$ obey their equations of motion.

Space of on-shell states of the $\mathcal{N}=4$ supemultiplet is naturally described
in a manifestly supersymmetric fashion by means of on-shell momentum superspace.
We are going to use its harmonic version:
\begin{eqnarray}
\mbox{$\mathcal{N}=4$ harmonic
on-shell momentum superspace}&=&\{\lambda_{\alpha},\tilde{\lambda}_{\dot{\alpha}},~\eta^{-}_{a},\eta^{+}_{a'},~u\}.
\end{eqnarray}
Here $\lambda_{\alpha},\tilde{\lambda}_{\dot{\alpha}}$ are the $SL(2,\mathbb{C})$
commuting spinors that parameterize
momenta carried by the on-shell state: $p_{\alpha\dot{\alpha}}=\lambda_{\alpha}\tilde{\lambda}_{\dot{\alpha}}$ if $p^2=0$.
All creation/annihlation operators of on-shell states, which are
two physical polarizations of gluons $|g^-\rangle, |g^+\rangle$,
four fermions $|\Gamma^A\rangle$ with positive and four fermions
$|\bar{\Gamma}^A\rangle$ with negative helicity, and three complex
scalars $|\phi^{AB}\rangle$ (anti-symmetric in the $SU(4)_R$ indices
$AB$ ) can be combined together into one $\mathcal{N}=4$ invariant
superstate ("superwave-function")
$|\Omega_{i}\rangle=\Omega_{i}|0\rangle$ ($i$ numerates momenta carried by the state):
\begin{eqnarray}\label{superstate}
|\Omega_{i}\rangle=\left(g^+_i + (\eta\Gamma_{i}) +
\frac{1}{2!}(\eta\eta\phi_{i}) +
\frac{1}{3!}(\varepsilon\eta\eta\eta\bar{\Gamma}_i) +
\frac{1}{4!}(\varepsilon\eta\eta\eta\eta )g^-_i\right)
|0\rangle,
\end{eqnarray}
where $(\ldots)$ represents contraction with respect to
the $SU(2) \times SU(2)' \times U(1)$
indices, $(\varepsilon\ldots)$ represents contraction
with $\varepsilon_{ABCD}$ symbol.
It is implemented, one has to express all $SU(4)$
indices in terms of $SU(2) \times SU(2)' \times U(1)$
once using the set of harmonic variables $u$.
The $n$ particle superstate
$|\Omega_n\rangle$ is then given by
$|\Omega_n\rangle=\prod_{i=1}^n\Omega_i|0\rangle$.
Note that on-shell momentum superspace is chiral.
Due to that and subtleties
\cite{HarmonyofFF_Brandhuber,BKV_SuperForm}
with on-shell realisation of the stress tensor supermultiplet in
 terms of the $W^{++}$ superfield, it is natural to consider the chiral
 (self dual) sector of the stress tensor supermultiplet only.
 This can be done by putting all $\bar{\theta}$ to 0 by
 hand in $T$ (this often called "chiral truncation"):
\begin{eqnarray}\label{superstate}
\mathcal{T}(x,\theta^+)=Tr(W^{++}W^{++})|_{\bar{\theta}=0}.
\end{eqnarray}
All operators from $\mathcal{T}$ are constructed of the fields of the self dual
part of the $\mathcal{N}=4$ supermultiplet.
Also, it is important to mention that all component
fields in $\mathcal{T}$ are off-shell.

So we can consider the form factors of chiral truncation
(self dual sector) of the $\mathcal{N}=4$
stress tensor supermultiplet $\mathcal{F}_n$:
\begin{eqnarray}
\mathcal{F}_n(\{\lambda,\tilde{\lambda},\eta\},x,\theta^{+})=
\langle\Omega_n|\mathcal{T}(x,\theta^{+})|0\rangle,
\end{eqnarray}
Here we are considering the colour
ordered object $\mathcal{F}_n$. The physical form factor
$\mathcal{F}_n^{phys.}$ in the planar limit\footnote{$g \rightarrow
0$ and $N_c \rightarrow \infty$ of $SU(N_c)$ gauge group so that
$\lambda=g^2N_c=$fixed.} should be obtained from $\mathcal{F}_n$ as:
\begin{eqnarray}
\mathcal{F}_n^{phys.}(\{\lambda,\tilde{\lambda},\eta\},x,\theta^{+})=\sum_{\sigma\in
S_n/Z_n}Tr(t^{a_{\sigma(1)}}\ldots
t^{a_{\sigma(n)}})\mathcal{F}_n(\sigma(\{\lambda,\tilde{\lambda},\eta\}),x,\theta^{+}),
\end{eqnarray}
where the sum runs over all possible none-cyclic permutations
$\sigma$ of the set $\{\lambda,\tilde{\lambda},\eta\}$ and the trace
involves $SU(N_c)$ $t^a$ generators in the fundamental
representation; the factor $(2\pi)^4g^{n-2}2^{n/2}$ is dropped.
The normalization $Tr(t^at^b)=1/2$ is used.

Let us now consider the general Grassmann structure of $\mathcal{F}_n$.
It is convenient to perform transformation from $\theta^+$ and $x$
to $q$ and the set of axillary variables $\{\lambda^{'}_{\alpha},\eta^{'-}_{a},
\lambda^{''}_{\alpha},\eta^{''-}_{a}\}$, $\lambda^{'}\eta^{'-}+\lambda^{''}\eta^{''-}=\gamma^{-}$:
\begin{equation}
\hat{T}[\ldots] = \int d^4x^{\alpha\dot{\alpha}}~d^{-4}\theta
\exp(iqx+\theta^{+}\gamma^{-})[\ldots],
\end{equation}
\begin{equation}\label{T[superFormfactor]}
Z_n (\{\lambda,\tilde{\lambda},\eta\},\{q,\gamma^{-}\}) =
\hat{T}[\mathcal{F}_n].
\end{equation}
Using supersymmetry arguments ($Z_n$ should be annihilated
by an appropriate set
of supercharges) one can say that in general
\cite{HarmonyofFF_Brandhuber,BKV_SuperForm}:
\begin{eqnarray}\label{T[superFormfactor]}
Z_n (\{\lambda,\tilde{\lambda},\eta\},\{q,\gamma^{-}\}) &=&
\delta^4(\sum_{i=1}^n\lambda_{\alpha}^i\tilde{\lambda}_{\dot{\alpha}}^i-q_{\alpha\dot{\alpha}})
\delta^{-4}(q^-_{a\alpha}+\gamma^-_{a\alpha})\delta^{+4}(q^+_{a'\alpha})
\mathcal{X}_n\left(\{\lambda,\tilde{\lambda},\eta\}\right),\nonumber\\
\mathcal{X}_n&=&\mathcal{X}_n^{(0)} + \mathcal{X}_n^{(4)} + \ldots +
\mathcal{X}_n^{(4n-8)},
\end{eqnarray}
where
\begin{eqnarray}
q^{+}_{a'\alpha}=\sum_{i=1}^n\lambda_{\alpha}^i\eta^{+}_{a'i},
~q^{-}_{a\alpha}=\sum_{i=1}^n\lambda_{\alpha}^i\eta^{-}_{ai}.
\end{eqnarray}
Grassmann delta functions are defined as
(see the appendix for the
whole set of definitions regarding Grassmann
delta functions and their integration)
\begin{eqnarray}
\delta^{\pm4}\left(q^{\pm}_{a'/a~\alpha}\right)=\prod_{a'/a,b'/b=1}^2
\epsilon^{\alpha\beta}q^{\pm}_{a'/a~\alpha}q_{b'/b~\beta}^{\pm}.
\end{eqnarray}
$\mathcal{X}^{(4m)}_n$ are the homogenous $SU(4)_R$ and $SU(2)\times SU(2)' \times U(1)$ invariant polynomials of the order of $4m$.
Hereafter, for saving space we will use the notation:
\begin{eqnarray}
\delta^{8}(q+\gamma)\equiv
\delta^{-4}(q^-_{a\alpha}+\gamma^-_{a\alpha})\delta^{+4}(q^+_{a'\alpha}).
\end{eqnarray}
Assigning helicity $\lambda=+1$ to $|\Omega_i\rangle$ and
$\lambda=+1/2$ to $\eta$ and $\lambda=-1/2$ to
$\theta$, one can see that $\mathcal{F}_n$ has overall
helicity $\lambda_{\Sigma}=n$, $\delta^{+4}$ has
$\lambda_{\Sigma}=2$, the exponential factor has $\lambda_{\Sigma}=0$. From this
we see that
$\mathcal{X}^{(0)}_n$ has $\lambda_{\Sigma}=n-2$,
$\mathcal{X}^{(4)}_n$ has $\lambda_{\Sigma}=n-4$, etc,
$\mathcal{X}^{(0)}_n$, $\mathcal{X}^{(4)}_n$ etc. are understood as
analogs \cite{DualConfInvForAmplitudesCorch} of the MHV, NMHV etc.
parts of the superamplitude, i.e., the part of the super form factor proportional
to the $\mathcal{X}^{(0)}_n$ will contain component form factors with
overall helicity $n-2$ which we will call the MHV form factors, part of
super form factor proportional to $\mathcal{X}^{(4)}_n$ will contain
component form factors with overall helicity $n-4$ which we will
call NMHV etc. up to $\mathcal{X}_n^{(4n-8)}$ overall helicity $2-n$
which we will call $\overline{\mbox{MHV}}$.

One can think \cite{HarmonyofFF_Brandhuber} that it is still possible
to describe the form factors of the full stress tensor supermultiplet
disregarding subtleties with on-shell realization,
at least at the tree level, using symmetry arguments and the full $W^{++}(x,\theta^+,\bar{\theta}^+)$
superfield. To do this, one has to introduce the
none chiral version of the on-shell momentum superspace,
which in our case can be obtained by performing the
following Grassmann Fourier transform:
\begin{eqnarray}
|\overline{\Omega}_{i}\rangle&=&\int d^{+2}\eta_i~exp(\eta^+_i\bar{\eta}^-_i)~|\Omega_{i}\rangle,\nonumber\\
|\overline{\Omega}_{i}\rangle&=&\left(g^+_i(\bar{\eta}^-_i\bar{\eta}^-_i )+... +
(\eta^-_i\eta^-_i )g^-_i\right)
|0\rangle.
\end{eqnarray}
After that one can define the form factor of full stress tensor supermultiplet
\begin{eqnarray}
\mathcal{F}_n^{full}(\{\lambda,\tilde{\lambda},\eta,\bar{\eta}\},x,\theta^{+},\bar{\theta}^+)=
\langle\overline{\Omega}_n|T(x,\theta^{+},\bar{\theta}^+)|0\rangle.
\end{eqnarray}
Performing $\hat{T}$ transformation from $(x,\theta^{+},\bar{\theta}^+)$ to $(q,\gamma^-,\bar{\gamma}^-)$ one can obtain $Z_n^{full} $:
\begin{eqnarray}\label{T[superFormfactor]}
Z_n^{full} (\{\lambda,\tilde{\lambda},\eta,\bar{\eta}\},\{q,\gamma^{-},\bar{\gamma}^{-}\}) &=&
\delta^4(\sum_{i=1}^n\lambda_{\alpha}^i\tilde{\lambda}_{\dot{\alpha}}^i-q_{\alpha\dot{\alpha}})
\delta^{-4}(q^-_{a\alpha}+\gamma^-_{a\alpha})
\delta^{-4}(\bar{q}^{-a'}_{\alpha}+\bar{\gamma}^{-a'}_{\alpha})\times\nonumber\\
&\times&\int \prod_{k=1}^n d^{+2}\eta_k~exp(\eta^+_k\bar{\eta}^-_k)
\delta^{+4}(q^+_{a'\alpha})
\mathcal{X}_n\left(\{\lambda,\tilde{\lambda},\eta\}\right),\nonumber\\
\end{eqnarray}
where now after Fourier transformation
\begin{eqnarray}\label{T[superFormfactor]}
\bar{q}^{-a'}_{\alpha}=\sum_{i=1}^n\lambda_{\alpha}^i\bar{\eta}^{-a'}.
\end{eqnarray}
We see that at least at the tree level the form factors of the full stress tensor
supermultiplet up to trivial Grassmann delta function are defined by the
Grassmann Fourier transformed $\mathcal{X}_n$ function, which one can compute
using chiral truncated (self dual sector) stress tensor supermultiplet only
\cite{HarmonyofFF_Brandhuber}.
Keeping this in mind we will focus on the self-dual sector form factors.

Using the BCFW recursion relations \cite{FormFactorMHV_component_Brandhuber}
one can show that for the MHV sector at the tree level one can obtain for $n$
point form factor (here we drop the momentum conservation delta function):
\begin{eqnarray}
Z_n^{(0)MHV}=\delta^{8}(q+\gamma)\mathcal{X}^{(0)}_n,~\mathcal{X}^{(0)}_n=\frac{1}{\langle 12\rangle\langle23\rangle...\langle n1\rangle}.
\end{eqnarray}
We will use this result in the next chapter.
Also, for completeness let us write down well known answers for tree level
$\mbox{MHV}_n$ and $\overline{\mbox{MHV}}_3$ amplitudes
\begin{equation}
A_n^{(0)MHV}=\frac{\delta^{8}(q)}{\langle 12\rangle\langle23\rangle...\langle
n1\rangle},~
A_3^{(0)\overline{MHV}}=\frac{\hat{\delta}^{4}(\eta_1[23]+\eta_2[31]+\eta_3[12])}{[12][23][31]},
\end{equation}
which will be used in the next section.

\section{BCFW and all-line shift for the NMHV sector}\label{p3}
Recursion relations for the tree level form factor were considered
in the literature before. BCFW recursion for the MHV sector, as was mentioned
earlier, was considered in \cite{FormFactorMHV_component_Brandhuber}
for the component form factors. All-line shift (CSW)
recursion for the NMHV sector was considered in \cite{HarmonyofFF_Brandhuber}
in the on-shell momentum superspace and momentum twistor spaces.
BCFW for form factors of more general $1/2$-BPS operators in the on
shell momentum superspace were considered in
\cite{FormFactorMHV_half_BPS_Brandhuber}.

In \cite{HarmonyofFF_Brandhuber}, it was argued that for the general
$[i,j \rangle$ shift the $\mbox{N}^k\mbox{MHV}$ form factor vanishes
as $z \rightarrow \infty$, so
BCFW recursion without "boundary terms".
Let us consider BCFW recursion for the NMHV sector in on-shell momentum superspace. Before
going to form factors it is useful to recall how BCFW recursion for the NMHV amplitudes works. It will help us to introduce important structures
and make useful analogies. For the adjacent $[i-1,i \rangle$ shift
\begin{eqnarray}
&&\hat{\lambda}_{i}=\lambda_{i}+z\lambda_{i-1},\nonumber\\
&&\hat{\tilde{\lambda}}_{i-1}=\tilde{\lambda}_{i-1}-z\tilde{\lambda}_{i},\nonumber\\
&&\hat{\eta}_i=\eta_i+z\eta_{i-1}.
\end{eqnarray}
there are two types of contributions in BCFW recursion in the NMHV sector,
which are combined of the $(\mbox{MHV}\otimes \mbox{MHV})$
and $(\mbox{NMHV}_{n-1}\otimes\overline{\mbox{MHV}}_3)$
amplitudes\footnote{$\otimes$ stands for summation over internal states (Grassmann integration) and substitution of the corresponding $z$ values.}.
The $\mbox{MHV}\otimes \mbox{MHV}$ terms are given
by so called $R_{rst}$ 2mh functions times the MHV tree level amplitude.
The $\mbox{NMHV}_{n-1}\otimes\overline{\mbox{MHV}}_3$ term can be represented
in terms of $R_{rst}$ functions as well. The $R_{rst}$ function can be written as:
\begin{figure}[t]
 \begin{center}
  \epsfxsize=7cm
 \epsffile{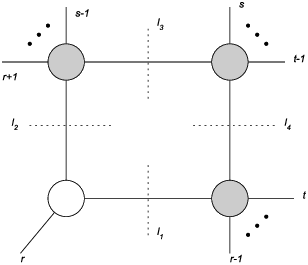}
 \end{center}\vspace{-0.2cm}
 \caption{Diagrammatic representation of the quadruple cut proportional to $R_{rst}$.
 The white blob is the $\overline{\mbox{MHV}}_3$ vertex and the light-grey blob is the MHV amplitude. }\label{NMHV_R_Amplitudes}
 \end{figure}
\begin{eqnarray}
R_{rst}=\frac{\langle ss-1\rangle\langle tt-1\rangle
\hat{\delta}^4(\Xi_{rst})}{x^2_{st}\langle r|x_{rt}x_{ts}|s\rangle\langle r|x_{rt}x_{ts}|s-1\rangle\langle r|x_{rs}x_{st}|t\rangle\langle r|x_{rs}x_{st}|t-1\rangle},
\end{eqnarray}
\begin{eqnarray}
\Xi_{rst}^A=\sum_{i=t}^{r-1}\eta_i^A\langle i|x_{ts}x_{sr}|r\rangle
+\sum_{i=r}^{s-1}\eta_i^A\langle i|x_{st}x_{tr}|r\rangle
=\langle \Theta_{tr}^A|x_{ts}x_{sr}|r\rangle+\langle \Theta_{rs}^A|x_{st}x_{tr}|r\rangle.
\end{eqnarray}
where $x_{ij}$ and $\Theta_{ij}^A$ are the dual variables defined as ($\langle l|\equiv\lambda_{l}$)
\begin{eqnarray}
x_{ij}=\sum_{k=i}^{j-1}p_k,~\langle \Theta_{ij}^A|=\sum_{l=i}^{j-1}\eta_l^A\langle l|.
\end{eqnarray}
In the harmonic superspace formulation
$\Xi_{rst}^A$ splits into $\Xi_{rst}^{+a}$
and $\Xi_{rst}^{-a'}$ as well as the Grassmann delta function
$\hat{\delta}^4=\hat{\delta}^{-2}\hat{\delta}^{+2}$ (see appendix for details).
Throughout the paper we will assume that numbers of momenta
$r,s,t,...$ ect. are
arranged anticlockwise for the form factors, were it is not mentioned
otherwise. All sums are understood in the
cyclic sense, for example, if n=6 s=5,t=3 then
$\sum_s^t=\sum_s^n+\sum_1^t=\sum_5^6+\sum_1^3$.
For $n\leq4$ the $R_{rst}$ function vanishes.
The $R_{rst}$ 2mh functions may also be
obtained by quadruple cuts of the one-loop
NMHV amplitude.
In fact, there is a deep connection between on-shell recursion relation
for tree level amplitudes and their loop level structure
\cite{Arcani_Hamed_PositiveGrassmannians,Masson_Skiner_Grassmaians_Twistors,BCFW_Tree,Generalized_unitarity_Cachzo}.
Also the $R_{rst}$ functions are invariants with respect to dual superconformal transformations
\cite{DualConfInvForAmplitudesCorch} from
dual $SU(2,2|4)$ as well as ordinary superconformal group $SU(2,2|4)$.
Even more, the $R_{rst}$ functions are invariants with respect to
full Yangian algebra \cite{Yangian_Drummond} which includes generators from dual and ordinary
superconformal algebras.
In harmonic superspace formulation the $R_{rst}$ functions are also invariants with respect to
$SU(2)\times SU(2)' \times U(1)$.
There is also interesting geometrical interpretation
\cite{Arcani_Hamed_Polytopes}
of them which we will discuss further in detail.
Using these functions one can write the results of
BCFW recursion for the NMHV sector for the amplitudes
for the $[1,2 \rangle$ shift as:
\begin{eqnarray}
A_n^{(0)NMHV}=\left(A_{n-1}^{(0)NMHV}\otimes A_3^{(0)\overline{MHV}}\right)+A_{n}^{(0)MHV}\sum_{i=4}^{n-1}R_{12i}.
\end{eqnarray}
This recursion relation can be solved in terms of the $R_{rst}$
functions (note that some terms in this sum
are actually equal to 0):
\begin{eqnarray}
A_n^{(0)NMHV}=A_{n}^{(0)MHV} \left(\sum_{j=2}^{n-2}\sum_{i=j+2}^{n}R_{1ji}\right).
\end{eqnarray}

It is natural to assume that the NMHV sector of the form factors can
be represented in terms of quadruple cut coefficients as in the case
of the amplitudes.
Quadruple cuts for the NMHV sector of the form factors were studied in \cite{BORK_NMHV_FF}.
There are three different types of analogs of the $R_{rst}$ functions
for the form factors $R^{(1)}_{rst}$, $R^{(2)}_{rst}$ and $\tilde{R}^{(1)}_{rtt}$ ("the $R$ functions").
\begin{figure}[t]
 \begin{center}
  \epsfxsize=7cm
 \epsffile{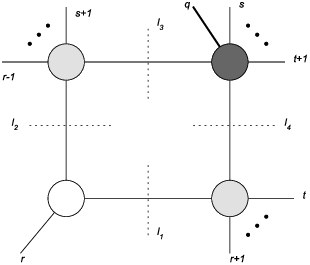}
 \end{center}\vspace{-0.2cm}
 \caption{Diagrammatic representation of the
 quadruple cut proportional to $R^{(1)}_{rst}$. 
 The dark grey blob is the MHV form factor.}\label{NMHV_R1}
 \end{figure}
\begin{figure}[t]
 \begin{center}
  \epsfxsize=7cm
 \epsffile{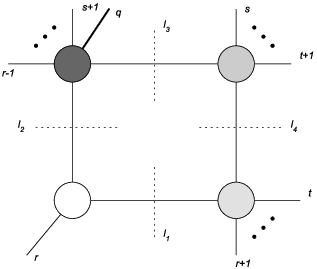}
 \end{center}\vspace{-0.2cm}
 \caption{Diagrammatic representation of the quadruple cut proportional to $R^{(2)}_{rst}$. }\label{NMHV_R2}
 \end{figure}
\begin{figure}[h]
 \begin{center}
  \epsfxsize=7cm
 \epsffile{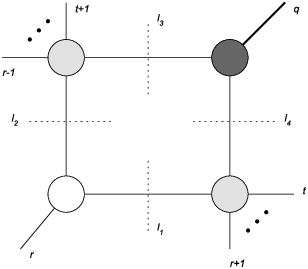}
 \end{center}\vspace{-0.2cm}
 \caption{Diagrammatic representation of the quadruple cut proportional to $\tilde{R}^{(1)}_{rtt}$. }\label{NMHV_R1lim}
 \end{figure}
\begin{eqnarray}\label{R_1_deff}
R_{rst}^{(1)}&=&\frac{\langle s+1s\rangle\langle
t+1t\rangle\hat{\delta}^4\left(\sum_{i=r+1}^{t}\eta_i\langle
i|p_{s+1...t}p_{s+1...r+1}|r\rangle-\sum_{i=r}^{s+1}\eta_i\langle
i|p_{s+1...t}p_{t...r+1}|r\rangle\right)} {p_{s+1...t}^2\langle
r|p_{r...s+1}p_{t...s+1}|t+1\rangle\langle
r|p_{r...s+1}p_{t...s+1}|t\rangle\langle
r|p_{t...r}p_{t...s+1}|s+1\rangle\langle
r|p_{t...r}p_{t...s+1}|s\rangle},\nonumber\\
\end{eqnarray}
\begin{eqnarray}\label{R_2_deff}
R_{rst}^{(2)}&=&\frac{\langle s+1s\rangle\langle
t+1t\rangle\hat{\delta}^4\left(\sum_{i=t}^{r+1}\eta_i\langle
i|p_{s...t+1}p_{r+1...s}|r\rangle+\sum_{i=r}^{s+1}\eta_i\langle
i|p_{s...t+1}p_{t...r+1}|r\rangle\right)} {p_{s...t+1}^2\langle
r|p_{r...s}p_{s...t+1}|t+1\rangle\langle
r|p_{r...s}p_{s...t-1}|t\rangle\langle
r|p_{t...r+1}p_{s...t+1}|s+1\rangle\langle
r|p_{t...r+1}p_{s...t+1}|s\rangle},\nonumber\\
\end{eqnarray}
\begin{eqnarray}\label{tildeR_1_deff}
\tilde{R}_{rtt}^{(1)}&=&\frac{\langle
tt+1\rangle\hat{\delta}^4\left(\sum_{i=t}^{r+1}\eta_i\langle
i|p_{1...n}p_{r...t+1}|r\rangle-\sum_{i=r}^{t+1}\eta_i\langle
i|p_{1...n}p_{t...r+1}|r\rangle\right)}{q^4\langle
r|p_{r...t+1}p_{1...n}|t\rangle\langle r|p_{t...r}q|t+1\rangle
\langle r|p_{t...r+1}p_{1...n}|r\rangle},\nonumber\\
\end{eqnarray}
where we used notations $p_{i_1...i_n}=p_{i_1}+...+p_{i_n}$,
$q=\sum_{i=1}^np_i$. The same notation will be used hereafter.
One can see that $R^{(1)}_{rst}$, $R^{(2)}_{rst}$ in fact
coincides with $R_{rst}$ computed in the corresponding kinematics,
while $\tilde{R}^{(1)}_{rtt}$ is different. Note also that due to the
presence of momenta $q$ carried by the operator in the momentum conservation condition
for the form factors $R^{(1)}_{rst}$, $R^{(2)}_{rst}$ and
$\tilde{R}^{(1)}_{rtt}$ can be defined (are none vanishing)
starting with the number of particles $n=3$.

Now we are ready to return to tree the level form factors. As it was stated
earlier we
hope that the NMHV sector of the form factors can be represented in terms of
the quadruple cut coefficients $R^{(1)}_{rst}$, $R^{(2)}_{rst}$ and
$\tilde{R}^{(1)}_{rtt}$. Indeed, this is just the case. By explicit
computation one can see that in the case of the $[1,2 \rangle$ shift:
\begin{eqnarray}
Z_n^{(0)NMHV}=\left(Z_{n-1}^{(0)NMHV}\otimes A_3^{(0)\overline{MHV}}\right)+Z_{n}^{(0)MHV}
\left(\tilde{R}^{(1)}_{122}+\sum_{i=3}^{n-1}R^{(1)}_{1i2}+\sum_{i=3}^{n}R^{(2)}_{1i2}\right).
\end{eqnarray}
Just as in the amplitude case the coefficients
$R^{(1)}_{1i2}$, $R^{(2)}_{1i2}$ and
$\tilde{R}^{(1)}_{122}$ are given by 2mh quadruple cuts.
Let us write several answers for some fixed $n$.
For example, for $n=3$ and $n=4$ one can get:
\begin{eqnarray}
Z_3^{(0)NMHV}=Z_{3}^{(0)MHV}\tilde{R}^{(1)}_{122},
\end{eqnarray}
\begin{figure}[t]
 \begin{center}
  \epsfxsize=11cm
 \epsffile{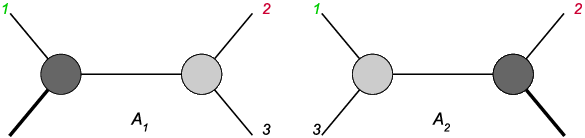}
 \end{center}\vspace{-0.2cm}
 \caption{BCFW diagrams contributing to the $n=3$ case,
 for the $[\textcolor{green}{1},\textcolor{red}{2} \rangle$ shift.
 $A_1=0$ due to the kinematic reasons.}\label{NMHV3}
 \end{figure}
\begin{eqnarray}
Z_4^{(0)NMHV}=Z_{4}^{(0)MHV}
\left(\tilde{R}^{(1)}_{133}+\tilde{R}^{(1)}_{122}+R^{(1)}_{132}+R^{(2)}_{142}\right).
\end{eqnarray}
Here $\tilde{R}^{(1)}_{133}$ is given by
$\left(Z_{3}^{(0)NMHV}\otimes A_3^{(0)\overline{MHV}}\right)$ term.
Note also that $R^{(2)}_{132}=0$ for the $n=4$ case.
In general the result for
$\left(Z_{n-1}^{(0)NMHV}\otimes A_3^{(0)\overline{MHV}}\right)$
for $Z_{n}^{(0)NMHV}$ can be conveniently written in terms of
$Z_{n-1}^{(0)NMHV}$ by introducing the shift operator
$\mathbb{S}$ that shifts the number of arguments
of the function starting with $2$ by $+1$,
(for example, $\mathbb{S}f(x_0,x_1,x_2,x_5)=f(x_0,x_1,x_3,x_6)$):
\begin{eqnarray}
\left(Z_{n-1}^{(0)NMHV}\otimes A_3^{(0)\overline{MHV}}\right)(\{\lambda_i,\tilde{\lambda}_i,\eta\}_{i=1}^n,\{q,\gamma\})=
Z_{n}^{(0)MHV}~\mathbb{S}\frac{Z_{n-1}^{(0)NMHV}}{Z_{n-1}^{(0)MHV}}(\{\lambda_i,\tilde{\lambda}_i,\eta\}_{i=1}^{n-1}).
\nonumber\\
\end{eqnarray}
$\{q,\gamma\}$ are unshifted. This can be seen using BCFW recursion
in the $\mbox{MHV}$
sector and representing $R$ functions as a quadruple cut that is
given by the product of the $\mbox{MHV}_n$ and $\overline{\mbox{MHV}}_3$
amplitudes and form factors
\footnote{In the amplitude case this can be most easily seen in the momentum twistor
formulation \cite{Henrietta_Amplitudes} or using of on-shell diagrams
\cite{Arcani_Hamed_PositiveGrassmannians}.
In the case of the form factors one may hope that the extension of on-shell
diagrams formalism also exists, but we are not going to discuss this issue here.}.
Using this observation one can write the answer for
$Z_{n}^{(0)NMHV}$ in closed form using
$R^{(1)}_{rst}$, $R^{(2)}_{rst}$ and $\tilde{R}^{(1)}_{rtt}$ functions:
\begin{eqnarray}
Z_{n}^{(0)NMHV}=Z_{n}^{(0)MHV}\left(\sum_{i=2}^{n-1}\tilde{R}^{(1)}_{1ii}+\sum_{i=2}^{n-2}\sum_{j=i+1}^{n-1}R^{(1)}_{1ji}+
\sum_{i=2}^{n-2}\sum_{j=i+2}^{n}R^{(2)}_{1ji}\right).
\end{eqnarray}
\begin{figure}[t]
 \begin{center}
  \epsfxsize=14cm
 \epsffile{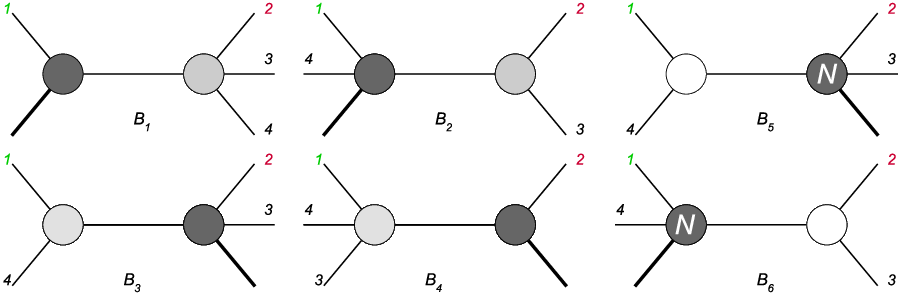}
 \end{center}\vspace{-0.2cm}
 \caption{BCFW diagrams contributing to the $n=4$ case, for the $[\textcolor{green}{1},\textcolor{red}{2} \rangle$ shift.
 $B_{2}=B_{5}=0$ due to the kinematic reasons.}\label{NMHV4_12}
 \end{figure}
 \begin{figure}[t]
 \begin{center}
  \epsfxsize=14cm
 \epsffile{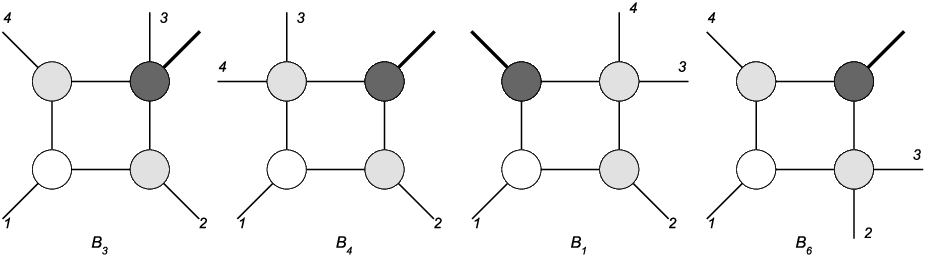}
 \end{center}\vspace{-0.2cm}
 \caption{Schematic representation of the corresponding $R$ functions contributing to the $n=4$ case, for the $[\textcolor{green}{1},\textcolor{red}{2} \rangle$ shift.}\label{NMHV4_12_R}
 \end{figure}
 \begin{figure}[h]
 \begin{center}
  \epsfxsize=14cm
 \epsffile{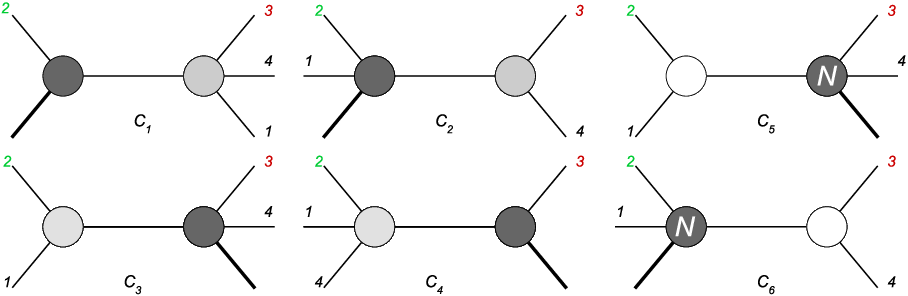}
 \end{center}\vspace{-0.2cm}
 \caption{BCFW diagrams contributing to the $n=4$ case, for the $[\textcolor{green}{2},\textcolor{red}{3} \rangle$ shift.
 $C_{2}=C_{5}=0$ due to the kinematic reasons.}\label{NMHV4_23}
 \end{figure}
 \begin{figure}[h]
 \begin{center}
  \epsfxsize=14cm
 \epsffile{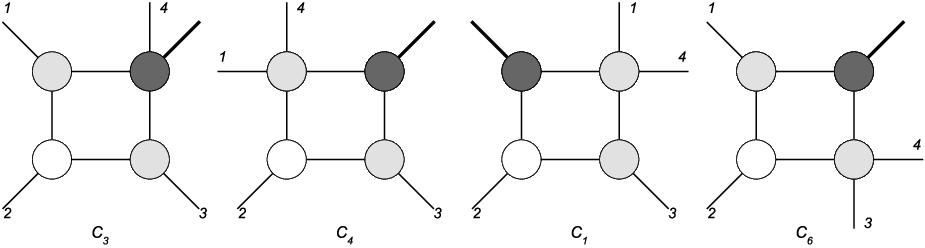}
 \end{center}\vspace{-0.2cm}
 \caption{Schematic representation of the corresponding $R$ functions contributing to the $n=4$ case, for the $[\textcolor{green}{2},\textcolor{red}{3} \rangle$ shift.}\label{NMHV4_12_R}
 \end{figure}

As a by product let us also consider different BCFW shifts. For example, for
the $[2,3 \rangle$ shift, $n=4$ one can obtain
the following representation of $Z_4^{(0)NMHV}$:
\begin{eqnarray}
Z_4^{(0)NMHV}=Z_{4}^{(0)MHV}
\left(\tilde{R}^{(1)}_{244}+R^{(1)}_{243}+R^{(2)}_{213}+\tilde{R}^{(1)}_{233}\right).
\end{eqnarray}
Adding the results of $[1,2 \rangle$ and $[2,3 \rangle$ shifts with the $1/2$ coefficient we obtain
representation of $Z_4^{(0)NMHV}$ computed as a coefficient of the IR pole at one loop
for the $\mbox{NMHV}$ form factor \cite{BORK_NMHV_FF}.
This can be written in the
following cyclic invariant
form\footnote{$\mathbb{P}$ is the permutation operator which shifts the number of all arguments of function by $+1$, i.e., for example: $\mathbb{P}f(x_0,x_1,x_2,x_5)=f(x_1,x_2,x_3,x_6)$.}
\cite{BORK_NMHV_FF}:
\begin{equation}
Z_4^{(0)NMHV} =Z_4^{(0)MHV}\frac{1}{2}
(1+\mathbb{P}+\mathbb{P}^2+\mathbb{P}^3)(\tilde{R}^{(1)}_{311}+R^{(1)}_{241}).
\end{equation}
Here we used the identity $R^{(2)}_{413}=R^{(1)}_{241}$ (see appendix).

As an illustration let us consider computation of the term which gives
$\tilde{R}^{(1)}_{122}$
in the $n=3$, $[1,2 \rangle$ shift case. For $n=3$ we have only one term
contributing to $Z_3^{(0)NMHV}=A_2$ which is given by
(see fig., $A_1=0$ due to the kinematic reasons):
\begin{eqnarray}
A_2=\int d^4\eta_{\hat{P}} Z_{2}^{(0)MHV}(\hat{2},\hat{P})\frac{1}{p^2_{13}}A^{(0)MHV}_{3}(-\hat{P},3,\hat{1}).
\end{eqnarray}
Preforming Grassmann integration and substituting 
$z_{13}=[13]/[23]$, $\hat{p}_{13}=p_{13}+z_{13}\lambda_1\tilde{\lambda}_2$ 
we obtain ($q_{123}=\lambda_1\eta_1+\lambda_2\eta_2+\lambda_3\eta_3$)
\begin{eqnarray}
A_2&=&\frac{\delta^8(q_{123}+\gamma)}{\langle 1\hat{p} \rangle \langle3\hat{p}\rangle \langle \hat{2}\hat{p}\rangle^2\langle 13\rangle}\int d^4\eta_{\hat{P}}\delta^8(\lambda_3\eta_3+\hat{\lambda}_1\hat{\eta}_1-\hat{\lambda}_P\hat{\eta}_P)\nonumber\\
&=&\frac{\delta^8(q_{123}+\gamma)\hat{\delta}^4\left( [2|\hat{p}_{13}|1\rangle\eta_1+[13]/[23][2|\hat{p}_{13}|1\rangle\eta_2+
[2|\hat{p}_{13}|3\rangle\eta_3 \right)}
{\langle 13 \rangle\langle1|\hat{p}_{13}|2]\langle3|\hat{p}_{13}|2]\langle\hat{2}|\hat{p}_{13}|2]^2p_{13}^2}
\nonumber\\
&=&\frac{\delta^8(q_{123}+\gamma)}{[12][23][31]}\frac{\hat{\delta}^4([23]\eta_1+[31]\eta_2+[21]\eta_3)}
{\langle \hat{2}|p_{13}|2]^2}.
\end{eqnarray}
After noting that
\begin{eqnarray}
\langle \hat{2}|p_{13}|2]=\langle 2|p_{13}|2]+p_{13}^2=q^2,
\end{eqnarray}
we can write (note also that momentum $q$ carried by the operator
is equal to $q=p_{123}$)
\begin{eqnarray}
A_2&=&\frac{\delta^8(q_{123}+\gamma)}{\langle12\rangle\langle23\rangle\langle31\rangle}\times
\frac{\langle12\rangle\langle23\rangle\langle31\rangle\hat{\delta}^4([23]\eta_1+[31]\eta_2+[21]\eta_3)}
{q^4[12][23][31]}=\nonumber\\
&=&\frac{\delta^8(q_{123}+\gamma)}{\langle12\rangle\langle23\rangle\langle31\rangle}\times
\frac{\langle23\rangle\hat{\delta}^4(\eta_2\langle2|qp_{13}|1\rangle
-\eta_{1}\langle1|qp_{21}|1\rangle
-\eta_{3}\langle3|qp_{21}|1\rangle)}
{q^4\langle1|p_{13}q|2\rangle\langle1|p_{12}q|3\rangle\langle1|p_{12}q|1\rangle}\nonumber\\&=&
Z_3^{(0)MHV}\times\tilde{R}^{(1)}_{122}.
\end{eqnarray}

The pole $q^4$ is canceled in this expression on the support of $\delta^8(q_{123}+\gamma)$. This will be important for us later on. Indeed $\delta^8(q+\gamma)=\delta^{-4}(q^-+\gamma^-)\delta^{+4}(q^+)$,
$q^{+}_{a'}=\sum_{i=1}^3\eta_{a',~i}^{+}\lambda_i$ so
\begin{eqnarray}
\hat{\delta}^{+2}\left(\eta_2^+\langle2|qp_{13}|1\rangle
-\eta_{1}^+\langle1|qp_{21}|1\rangle
-\eta_{3}^+\langle3|qp_{21}|1\rangle\right)=\nonumber\\
\hat{\delta}^{+2}\left(\eta_2^+\langle21\rangle q^2-\sum_{i=1}^3\eta_i^+\langle i|qp_{12}|1\rangle\right)=\nonumber\\
\hat{\delta}^{+2}\left(\eta_2^+\langle21\rangle q^2-0\right)=q^4\langle21\rangle^2\hat{\delta}^{+2}\left(\eta_2^+\right).
\end{eqnarray}
Cancellation of the $q^4$ pole will be true also for arbitrary $n$ for
$\tilde{R}^{(1)}_{rtt}$.

Let us briefly discuss analytical properties of the results of BCFW recursion.
As an example
we will consider $n=4$ case. Each $R^{(1)}_{rst},R^{(2)}_{rst}$
and $\tilde{R}^{(1)}_{rtt}$ term is a rational function of
$\lambda_i,\tilde{\lambda}_i$ variables and has several poles.
Some of them are physical, i.e., correspond to appropriate factorisation channels
\cite{BCFW_Tree}, while others are spurious
and must be canceled in the whole sum.
The presence of spurious poles is the general feature of BCFW recursion,
and its application to the form factors is no exception.
So in the $n=4$, $[1,2 \rangle$ shift case the list of poles is the following:
\begin{eqnarray}
R^{(1)}_{132}:~\langle3|q|2],\langle3|q|4],~p_{124}^2,p_{12}^2,p_{14}^2,~
R^{(2)}_{142}:~\langle1|q|4],\langle1|q|2],~p_{234}^2,p_{34}^2,p_{23}^2,
\end{eqnarray}
\begin{eqnarray}
\tilde{R}^{(1)}_{122}:~\langle3|q|2],\langle1|q|2],~p_{134}^2,~
\tilde{R}^{(1)}_{133}:~\langle1|q|4],\langle3|q|4],~p_{123}^2.
\end{eqnarray}
Poles
\begin{eqnarray}
p_{123}^2,p_{124}^2,p_{234}^2,p_{123}^2,p_{12}^2,p_{23}^2,p_{34}^2,p_{41}^2,
\end{eqnarray}
are physical, while
\begin{eqnarray}
\langle1|q|2],\langle1|q|4],\langle3|q|2],\langle3|q|4],
\end{eqnarray}
are spurious once. The structure of $Z^{(0)NMHV}_4$ suggests
that spurious poles should cancel themselves between
the $R$ functions (for example $\langle1|q|2]$ should be canceled
between $R^{(2)}_{142}$ and $\tilde{R}^{(1)}_{122}$) but it is
not easy to see how it really works. Also it would be nice to
observe some general pattern of such cancelations for general $n$.
There are also several related questions.

1.One can consider a different type of recursion relations for the form factors:
all-line shift (CSW) \cite{CSW_Tree,Elvang_All_Line_Shift_1,Elvang_All_Line_Shift_2}.
Indeed, one can show that under anti holomorphic all-line
shift the form factors with operators from the stress-tensor
supermultiplet (number of fields in operator $m=2$,) behave as:
\begin{equation}
Z_n(z)\to z^{s} \mbox{(or better) as}~z\to \infty,~\mbox{with}~s=\frac{2-n+\lambda_{\Sigma}}{2},
\end{equation}
(note $2-n$ instead of $n-4$ \cite{Elvang_General_Field} in the amplitude
case due to the different mass dimension of the form factor),
so for $\lambda_{\Sigma}=n-4$, as in the $\mbox{NMHV}$ case recursion
is valid.
Thus one can easily obtain \cite{FormFactorMHV_half_BPS_Brandhuber}:
\begin{eqnarray}
Z_n^{(0)NMHV}=Z_{n}^{(0)MHV}\left(\sum_{i=1}^n\sum_{j=i+2}^{i+1-n}R_{*ij}\right),
~\mbox{with}~\lambda^{*}=0,\eta_{*}^A=0.
\end{eqnarray}
Here we exchange the problem of cancellation of
spurious poles to the problem of proving that
the poles of the form $\langle i|q|*]$ should be canceled. This
cancellation will imply that the result
is independent of the choice of $\tilde{\lambda}^*$
\cite{Arcani_Hamed_Polytopes}.
Note also that representations for $\mbox{NMHV}$ sector given by BCFW and
all-line shift (CSW) recursions naively look rather different.
It would be nice to show how one can transform one into another.

2.It would also be nice to write some simple recursion relation for
the general $\mbox{N}^k\mbox{MHV}$
form factor.

3.In the one loop generalised unitarity based computations
(for example, see \cite{BORK_NMHV_FF})
one encounters different none obvious relations between
$R$ functions. It would be nice to have some simple representation
for $R$ functions where these relations becomes obvious.

These questions are not unique to the form factors and one encounters
their analogs in the amplitude case as well.
In the case of amplitudes they all can be answered in beautiful
geometrical picture based on the momentum twistor representation
and the interpretation of the amplitudes as the volumes of polytopes
in $\mathbb{C}\mathbb{P}^4$ in the first non trivial $\mbox{NMHV}$ case
\cite{Hoges_Polytopes,Masson_Skiner_Grassmaians_Twistors,Arcani_Hamed_Polytopes}
and more general "Amplituhidron" picture
\cite{Arcani_Hamed_Amplituhdron_1,Arcani_Hamed_Amplituhdron_2} based
on positive Grassmanian geometry
\cite{Arcani_Hamed_PositiveGrassmannians} in the general case.

We are going to show now that in the case of the form factors one can also
use nearly the same momentum twistor representation to answer all
these questions.
Only one new ingredient is necessary - infinite periodical contour
in the momentum twistor space \cite{HarmonyofFF_Brandhuber}.

\section{Momentum twistor space representation}\label{p4}
To use momentum twistors, one has to introduce dual
variables $x_i$ for momenta $p_i$ \cite{Hoges_Polytopes}.
\begin{eqnarray}
p_{i}^{\alpha\dot{\alpha}}=x^{\alpha\dot{\alpha}}_{i}-x^{\alpha\dot{\alpha}}_{i-1}.
\end{eqnarray}
and their fermionic counterparts $q^{-}_{a\alpha, i}=\lambda_{\alpha, i}\eta^{-}_{ai}$, $q^{+}_{a'\alpha, i}=\lambda_{\alpha, i}\eta^{+}_{a'i}$ and $\Theta^{-}_{a\alpha}$, $\Theta^{+}_{a'\alpha}$:
\begin{eqnarray}
q^{-}_{a\alpha, i}=\Theta^{-}_{a\alpha, i}-\Theta^{-}_{a\alpha, i-1},
\end{eqnarray}
\begin{eqnarray}
q^{+}_{a'\alpha, i}=\Theta^{+}_{a'\alpha, i}-\Theta^{+}_{a'\alpha, i-1}.
\end{eqnarray}
This is where periodical configuration first appears
\cite{FormFactorMHV_component_Brandhuber,HarmonyofFF_Brandhuber}.
Indeed, we are working with a colour ordered object, so positions of momenta
of external particles $p_i$ are fixed. But the operator, which carries
the momentum
$q$, is colour singlet and can be inserted between any pair of momenta.
The same is true also for the fermionic counterpart of $q$, when
we are dealing with the
superspace formulation of the form factors.
One can think of working with different (with respect to position
where $q$ is inserted) closed contours, but it is not obvious how
to combine terms defined on different contours.
An infinite periodical (with period equal to $q$) configuration solves this problem.
In fact we will need only $2n$ $x_i$ independent variables to describe any
kinematic invariant $p_{1_1,...,i_l}^2$ we may encounter in the case of
$n$ particle form factor, at least at the tree level in the MHV and NMHV sectors.
The only feature that the periodical contour brings into play
and one should take into account is some sort of redundancy.
Everything is defined up to the shift over $k$ periods along the contour,
so one should
"gauge fix" which periods will be used.
Also the periodical configuration is very natural from the $AdS/CFT$ point of view
\cite{Zhiboedov_Strong_coupling_FF}. The insertion of operator corresponds
to consideration of a closed
string state on the string worldsheet in addition to open ones
(which correspond to particles) in the dual picture.
After such insertion, $T$-duality transformation gives infinite periodical
configuration with a period equal to momenta carried by the closed string state.
The periodical contour and hence dual variables $\Theta^-,\Theta^+$
can also be introduced to the total super momentum carried by particles
$q^{+},q^{-}$. The period will be equal to the super momenta
$\gamma^{+},\gamma^{-}$ carried by the operator.
Note that since $\gamma^{+}=0$,
the corresponding fermionic part of the contour in the superspace will be closed.
\begin{figure}[t]
 \begin{center}
  \epsfxsize=14cm
 \epsffile{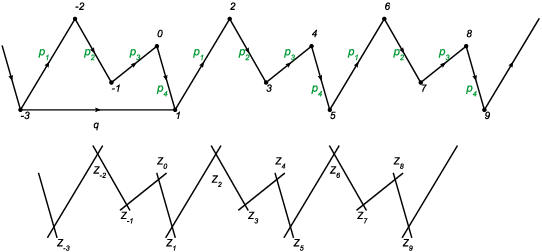}
 \end{center}\vspace{-0.2cm}
 \caption{Dual contour in momentum and momentum twistor spaces for $n=4$ form factor.}\label{NMHV_R1}
 \end{figure}

Now we are ready to introduce momentum supertwistors
\cite{Hoges_Polytopes,Masson_Skiner_Grassmaians_Twistors}.
The points in the dual superspace are mapped to the lines in momentum twistor
space $(x_i,\Theta_{i})\sim\mathcal{Z}_{i-1}\wedge\mathcal{Z}_{i}$
(as usual $i$ is the number of a particle, with:
\begin{eqnarray}
\mathcal{Z}_i^{\pm\Delta}=\left(
\begin{array}{ccc}
  Z_{i}^M \\
  \chi_{a/a',i}^{\pm}
\end{array}
\right),
\end{eqnarray}
The fermionic part of the supertwistor $\chi$ is given by:
\begin{eqnarray}
\chi_{ai}^{-}=\Theta_{ai}^-\lambda_{i},~\chi_{a'i}^{+}=\Theta_{ai}^+\lambda_{i}.
\end{eqnarray}
Note that $\chi_{ai}^{-}$ part of the supertwistor belongs to
the infinite periodical contour,
$\chi_{ai}^{+}$ belongs to the  "closed part of the fermionic contour"
due to the $\gamma^{+}=0$ condition. Sometimes it will be convenient
to consider $\chi_{ai}^{+}$ also as part of the infinite periodical contour
in the intermediate expressions
and apply $\gamma^{+}=0$ only at the end.
Since all our expressions are polynomials in Grassamann
variables, a smooth limit in $\gamma^{+} \to 0$ always exists.
The bosonic part of the supertwistor is
\begin{eqnarray}
Z_i^{M}=\left(
\begin{array}{ccc}
  \lambda_{i}^{\alpha} \\
  \mu^{\dot{\alpha}}_i
\end{array}
\right),~
\mu^{\dot{\alpha}}_i=x^{\alpha\dot{\alpha}}_i\lambda_{\alpha i},
\end{eqnarray}
where $M=1...4$. The corresponding objects transform
under the action of
 the dual conformal $SU(2,2)$ group;
$\Delta$ is the multi-index for the $SU(2,2)$ indices and $\pm a/ \pm a'$
indices of $SU(2)$
and $U(1)$.
The standard notation for dual conformal $SU(2,2)$ invariant will be used:
\begin{eqnarray}
\langle i,j,k,l\rangle=\epsilon_{ABCD}Z^A_iZ^B_jZ^C_kZ^D_l.
\end{eqnarray}
In terms of the components of the twistors this expression can be written as
(here $\epsilon_{\dot{\alpha}\dot{\beta}}\mu_i^{\dot{\alpha}}\mu_j^{\dot{\beta}}\doteq[ij]$):
\begin{eqnarray}
\langle i,j,k,l\rangle=\langle ij\rangle[kl]+\langle ik\rangle[lj]+
\langle il\rangle[jk]+\langle kl\rangle[ij]+
\langle lj\rangle[ik]
+\langle jk\rangle[il].
\end{eqnarray}
Hereafter we will drop indices on the twistors and their
components everywhere when it does not lead to misunderstanding.
Due to the periodical configuration with period the
$q=\sum_{i=1}^{n}p_i$ we will have the following
relation for the momentum twistors:
\begin{eqnarray}
Z_{i}=(\lambda_{i},x_{i}\lambda_{i}),~
Z_{i+nk}=(\lambda_{i},x_{i+nk}\lambda_{i}),~i=1...n,~k\in\mathbb{N}.
\end{eqnarray}

Using $\langle i,j,k,l\rangle$ one can write the
following expressions for kinematical invariants
and products of spinors:
\begin{eqnarray}
\left(\sum_{l=i}^{j-1}p_l\right)^2=x_{ij}^2=\langle ii+1\rangle
\langle jj+1\rangle \langle i,i+1,j,j+1\rangle,
\end{eqnarray}
and
\begin{eqnarray}
\langle tt+1\rangle \langle r|x_{rt}x_{ts}|s\rangle = \langle r, t,t+1, s\rangle.
\end{eqnarray}
Because we are working with the periodical configuration,   one can
shift simultaneously all numbers
in $\langle r, t+1,t, s\rangle$ and $\langle i,i+1,k,k+1\rangle$
by $kn,~k\in\mathbb{N}$ without
changing the result. In addition there are several relations
between $\langle a,b,c,d\rangle$ invariants unique to the periodical
contour. We will need two of them:
\begin{eqnarray}\label{Relation_<abcd>_1}
\langle 1, i,i+1, 1+n\rangle=\langle 1, i-n,i+1-n, 1-n\rangle,
\end{eqnarray}
and
\begin{eqnarray}\label{Relation_<abcd>_2}
\langle i+n,i+1+n,i+2+n,i+1\rangle=\langle i,i+1,i+2,i+1+n\rangle.
\end{eqnarray}

As it was claimed before, in the case of the amplitudes
(closed contour) the $R_{rst}$ function is
invariant with respect to the dual superconformal transformations from
$SU(2,2|4)$. Using momentum the supertwistors one can see that
the following combination of 5 arbitrary twistors is
$SU(2,2|4)$ invariant \cite{Masson_Skiner_Grassmaians_Twistors}:
\begin{eqnarray}
[a,b,c,d,e]=\frac{\hat{\delta}^4(\langle a,b,c,d\rangle\chi_e+\mbox{cycl.})}
{\langle a,b,c,d\rangle\langle b,c,d,e\rangle\langle c,d,e,a\rangle\langle d,e,a,b\rangle
\langle e,a,b,c\rangle}.
\end{eqnarray}
Here $\hat{\delta}^4=\hat{\delta}^{-2}\hat{\delta}^{+2}$ (see the appendix).
The $R_{rst}$ function is a special case of this invariant:
\begin{eqnarray}
R_{rst}=[r,s,s+1,t,t+1].
\end{eqnarray}
\begin{figure}[t]
 \begin{center}
  \epsfxsize=12cm
 \epsffile{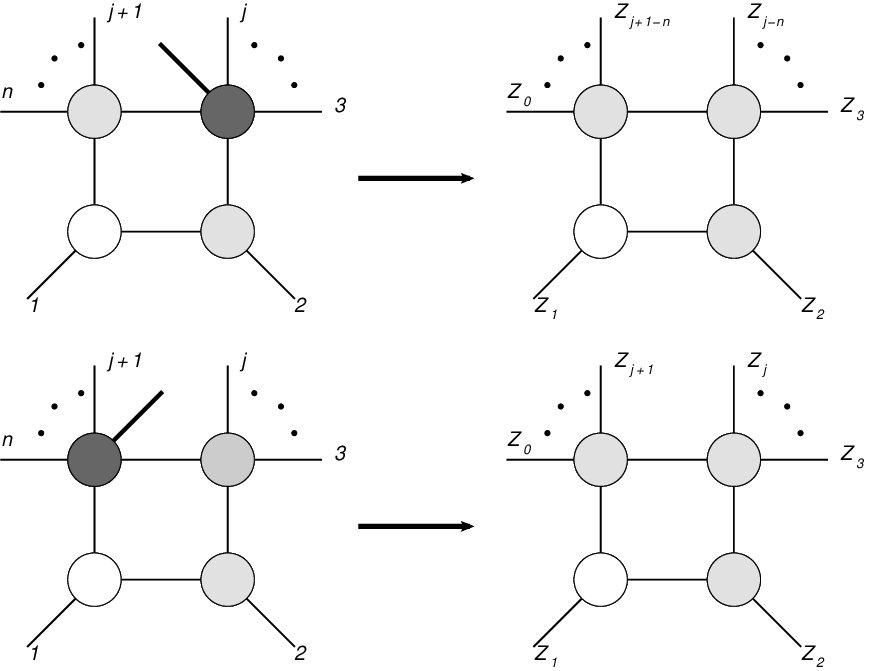}
 \end{center}\vspace{-0.2cm}
 \caption{Schematic representation of the relation between $R^{(1)}_{rst}$, $R^{(2)}_{rst}$
 functions and the corresponding $[abcde]$ momentum
 twistor dual superconformal invariants in the 2mh case.}\label{R1}
 \end{figure}
What about the $R^{(1)}_{rst}$, $R^{(2)}_{rst}$ and
$\tilde{R}^{(1)}_{rtt}$ functions for the form factors ? Using the momentum
supertwistors defined on periodical contour one can see that the
following identities hold for $R^{(1)}_{1st}$, $R^{(2)}_{1st}$:
\begin{eqnarray}
R^{(1)}_{1st}=[1,t,t+1,s-n,s+1-n],
\end{eqnarray}
and
\begin{eqnarray}
R^{(2)}_{1st}=[1,t,t+1,s,s+1].
\end{eqnarray}
Here $n$ is the number of twistors (particles) in period of the contour.
As it was explained earlier, due to the periodical nature of momentum twistor
configuration we are considering, this form is not unique.
For example, for $n=4$ one can see that:
$
[5,6,7,3,4]=[1,2,3,-1,0].
$
We choose this particular form ("fix the gauge") because it naturally
arises in the $[1,2\rangle$ shift.
\emph{It is implemented that the condition $\gamma^{+} = 0$
is imposed in the argument of
$\hat{\delta}^4$}.
The case of $\tilde{R}^{(1)}_{rtt}$ is special.
Nevertheless, it is also possible to rewrite
it in terms of the $[a,b,c,d,e]$ momentum twistor invariant
but with the nontrivial "bosonic" coefficient:
\begin{eqnarray}
\tilde{R}^{(1)}_{1tt}&=&c^{(n)}_t[1,t,t+1,t-n,t+1-n],\nonumber\\
c^{(n)}_t&=&\frac{\langle 1,t,t+1,t-n\rangle\langle 1,t-n,t+1-n,t+1\rangle}
{\langle1,t,t+1,1+n\rangle\langle t,t+1,t-n,t+1-n\rangle}.
\end{eqnarray}
\begin{figure}[t]
 \begin{center}
  \epsfxsize=12cm
 \epsffile{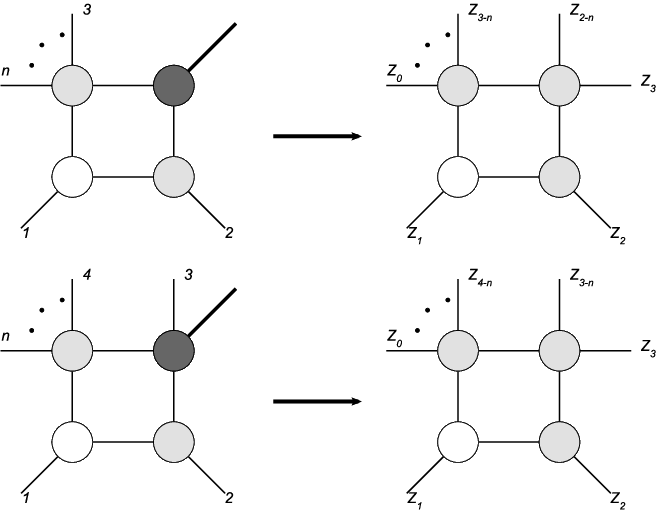}
 \end{center}\vspace{-0.2cm}
 \caption{Schematic representation of relation between $R^{(1)}_{rst}$
 functions and corresponding $[abcde]$ momentum twistor dual superconformal invariants, special cases.}\label{R1}
 \end{figure}
 \begin{figure}[t]
 \begin{center}
  \epsfxsize=12cm
 \epsffile{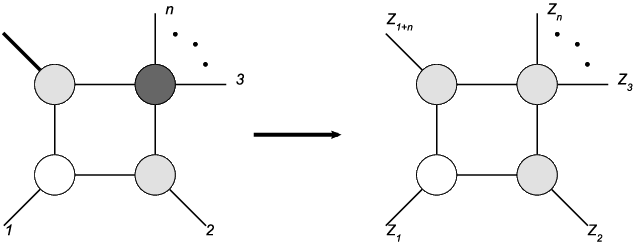}
 \end{center}\vspace{-0.2cm}
 \caption{Schematic representation of the relation between the $R^{(2)}_{rst}$
 functions and the corresponding $[abcde]$ momentum twistor dual superconformal invariants, special case.}\label{R1}
 \end{figure}
As an illustration how one can rewrite the $R$ coefficients in the momentum
twistor variables, let us consider the $n=4$ case (as usual we have
$q=p_{1234}$), $\tilde{R}^{(1)}_{122}$:
\begin{eqnarray}
\tilde{R}^{(1)}_{122}=\frac{\langle 23\rangle \hat{\delta}^4(X_{122})}{q^4\langle1|p_{12}q|3\rangle\langle1|p_{34}q|2\rangle\langle1|p_{34}q|1\rangle},
\end{eqnarray}
where
\begin{eqnarray}
X_{122}&=&-\eta_2\langle2|qp_{134}|1\rangle+\sum_{i=1,3,4}\eta_i\langle i|qp_{2}|1\rangle=\nonumber\\
&=&-\sum_{i=1,2}\eta_i\langle i|qp_{34}|1\rangle+\sum_{k=3,4}\eta_k\langle k|qp_{12}|1\rangle.
\end{eqnarray}
Now using the momentum twistors we can write:
\begin{eqnarray}
\langle -1,-2,2,3\rangle=\langle23\rangle\langle-1|x_{-13}x_{2-2}|-2\rangle=
\langle23\rangle\langle3|q(q+2)|2\rangle=\langle23\rangle^2q^2,
\end{eqnarray}
\begin{eqnarray}
\langle 1,-1,-2,2\rangle=\langle23\rangle\langle1|x_{1-1}x_{-12}|2\rangle=
\langle23\rangle\langle1|p_{34}q|2\rangle,
\end{eqnarray}
\begin{eqnarray}
\langle 1,2,3,-1\rangle=\langle23\rangle\langle1|x_{13}x_{3-1}|1\rangle=
\langle23\rangle\langle1|p_{12}q|3\rangle,
\end{eqnarray}
\begin{eqnarray}
\langle 1,-1,-2,-3\rangle=\langle-1-2\rangle\langle1|x_{1-1}x_{-1-3}|-3\rangle=
\langle23\rangle\langle1|p_{34}q|1\rangle.
\end{eqnarray}
So substituting this relations in $\tilde{R}^{(1)}_{122}$ one can see that:
\begin{eqnarray}
\tilde{R}^{(1)}_{122}&=&
\frac{\langle 23\rangle^8 \hat{\delta}^4(X_{122})}
{\langle-1,-2,2,3\rangle^2\langle1,-1,-2,2\rangle\langle1,-1,2,3\rangle\langle1,-1,-2,-3\rangle}
=
\nonumber\\
&=&
\frac{\langle1,2,3,-2\rangle\langle1,-2,-1,3\rangle}
{\langle1,-2,-1,-3\rangle\langle2,3,-2,-1\rangle}\times\nonumber\\
&\times&
\frac{\langle 23\rangle^8 \hat{\delta}^4(X_{122})}
{\langle1,-1,-2,2\rangle\langle-1,-2,2,3\rangle\langle-2,2,3,1\rangle\langle2,3,1,-1\rangle\langle3,1,-1,-2\rangle}.
\nonumber\\
\end{eqnarray}
From the last expression one can conclude that (we used (\ref{Relation_<abcd>_1}), which in
this case gives us $\langle1,2,3,4\rangle=\langle1,-2,-1,-3\rangle$)
\begin{eqnarray}
\frac{\langle1,2,3,-2\rangle\langle1,-2,-1,3\rangle}
{\langle1,-2,-1,-3\rangle\langle2,3,-2,-1\rangle}=c^{(4)}_2.
\end{eqnarray}
Now let us rewrite $X_{122}$ in terms of the momentum supertwistors
(here we suppress
$SU(2)\times SU(2)'\times U(1)$ indices, $\langle\Theta_{ij}|\equiv\Theta_{ij}$,
$\langle i|\equiv\lambda_i$). Here we treat $\chi^+_i$ and $\chi^-_i$
one equal footing, and will take the $\gamma^+ \to 0$ limit only in the
final expression.
One can see that on the periodical contour
\begin{eqnarray}
\langle\Theta_{13}|=\sum_{i=1,2}\eta_i\langle i|,~
\langle\Theta_{-11}|=-\sum_{i=3,4}\eta_i\langle i|,
\end{eqnarray}
and
\begin{eqnarray}
x_{-11}=p_{34},~x_{-13}=-x_{3-1}=q,~x_{31}=-p_{12},
\end{eqnarray}
so
\begin{eqnarray}
X_{133}&=&\langle\Theta_{13}|x_{3-1}x_{-11}|1\rangle+\langle\Theta_{1-1}|x_{-13}x_{31}|1\rangle.
\end{eqnarray}
Then we can write \cite{Masson_Skiner_Grassmaians_Twistors}:
\begin{eqnarray}
\langle\Theta_{13}|x_{3-1}x_{-11}|1\rangle+\langle\Theta_{1-1}|x_{-13}x_{31}|1\rangle=
\frac{\chi_1\langle2,3,-1,-2\rangle+perm.}{\langle23\rangle^2}.
\end{eqnarray}
Substituting this in $\tilde{R}^{(1)}_{122}$ we get:
\begin{eqnarray}
\tilde{R}^{(1)}_{122}&=&c_{2}^{(4)}
\frac{\langle 23\rangle^8 \hat{\delta}^4(X_{122})}
{\langle1,-1,-2,2\rangle\langle-1,-2,2,3\rangle\langle-2,2,3,1\rangle\langle2,3,1,-1\rangle\langle3,1,-1,-2\rangle}
\nonumber\\
&=&c_{2}^{(4)}[1,2,3,-2,-1],
\end{eqnarray}
as expected. Also, now we can take the $\gamma^+ \to 0$ limit.

Using these results one can easily rewrite
the BCFW recursion relations in the $\mbox{NMHV}$
sector for the form factors in momentum supertwistors
(hereafter we drop the $(0)$ subscript for simplicity):
\begin{eqnarray}
&&\frac{Z_n^{NMHV}(\mathcal{Z}_{2-n},...,\mathcal{Z}_{1+n})}{Z_{n}^{MHV}}=
\frac{Z_{n-1}^{NMHV}}{Z_{n-1}^{MHV}}(\mathcal{Z}_{2-n},...,\mathcal{Z}_{1},\mathcal{Z}_{3},\mathcal{Z}_{4},...,\mathcal{Z}_{1+n})+
\nonumber\\
&+&\sum_{j=3}^n[1,2,3,j,j+1]+\sum_{j=3}^{n-1}[1,2,3,j-n,j+1-n]+
c^{(n)}_2[1,2,3,2-n,3-n].\nonumber\\
\end{eqnarray}
As an illustration let us write the answers for
$n=3,4,5$ in the momentum supertwistor notations:
\begin{eqnarray}
\frac{Z_3^{NMHV}}{Z_{3}^{MHV}}=c^{(3)}_2[-1,0,1,2,3],
\end{eqnarray}
\begin{eqnarray}
\frac{Z_4^{NMHV}}{Z_{4}^{MHV}}&=&\left(\mathbb{S}c^{(3)}_2\right)[-1,0,1,3,4]+[1,2,3,4,5]
+[1,2,3,0,-1] \nonumber\\&+&c^{(4)}_2[1,2,3,-2,-1],
\end{eqnarray}
\begin{eqnarray}
\frac{Z_5^{NMHV}}{Z_{5}^{MHV}}
&=&\left(\mathbb{S}^2c^{(3)}_2\right)[-1,0,1,4,5]+[1,3,4,5,6]+[-1,0,1,3,4]
\nonumber\\&+&\left(\mathbb{S}c^{(4)}_2\right)[-2,-1,1,3,4]+
[1,2,3,4,5]+[1,2,3,5,6]\nonumber\\
&+&[1,2,3,-2,-1]+[1,2,3,-1,0]+c^{(5)}_2[1,2,3,-3,-2].
\end{eqnarray}

As a by product using these explicit expressions let us discuss
the relation between the form factors with the
supermomentum carried by the operator equal to zero and the amplitudes.
In \cite{BKV_SuperForm,HarmonyofFF_Brandhuber,Sasha}
it was observed that the following
relation between the form factors and amplitudes likely holds
\begin{equation}\label{cojectureAmpl-FF}
Z_n(\{\lambda,\tilde{\lambda},\eta\},\{0,0\})= g\frac{\partial
A_n(\{\lambda,\tilde{\lambda},\eta\})}{\partial g}.
\end{equation}

In our momentum supertwistor notation the limit of
$q\rightarrow 0$, $\gamma^{\pm}\rightarrow0$
corresponds to "gluing" all periods of the contour together,
i.e., for the $n$ particle case
$\mathcal{Z}_i\rightarrow\mathcal{Z}_{i\pm kn}$
for any integer $k$ and $i$. Taking
this limit in written above answers for the form factors
one can see that (remember that
$[a,b,c,d,e]=0$ if any of the two arguments coincide):
\begin{eqnarray}
Z_3^{NMHV}|_{\mathcal{Z}_i\rightarrow\mathcal{Z}_{i\pm k3}}=0,
\end{eqnarray}
\begin{eqnarray}
Z_4^{NMHV}|_{\mathcal{Z}_i\rightarrow\mathcal{Z}_{i\pm k4}}=0,
\end{eqnarray}
\begin{eqnarray}
Z_5^{NMHV}|_{\mathcal{Z}_i\rightarrow\mathcal{Z}_{i\pm k5}}\sim A_5^{MHV}[1,2,3,4,5]=A_5^{NMHV},
\end{eqnarray}
as one would expect because there are no $3$ and $4$ point $\mbox{NMHV}$
amplitudes.\footnote{Actually we obtained $Z_5^{NMHV}|_{\mathcal{Z}_i\rightarrow\mathcal{Z}_{i\pm k5}}=2 A_5^{MHV}[1,2,3,4,5]$. The presence of coefficient $2$ is unexpected. However one can
also see (?) that from the BCFW representation of the $\mbox{N}^k\mbox{MHV}$ form factors
that the coefficient will be $k+1$ in the $\mbox{N}^k\mbox{MHV}$ sector.}
Note also that in our case of the super form factors, this limit
is well defined and can be easily taken, while in components it is singular
for some particular answers and in on-sell momentum superspace
\cite{BKV_SuperForm}
it is not obvious at first glance how exactly these singularities are canceled.
\begin{figure}[t]
 \begin{center}
  \epsfxsize=17cm
 \epsffile{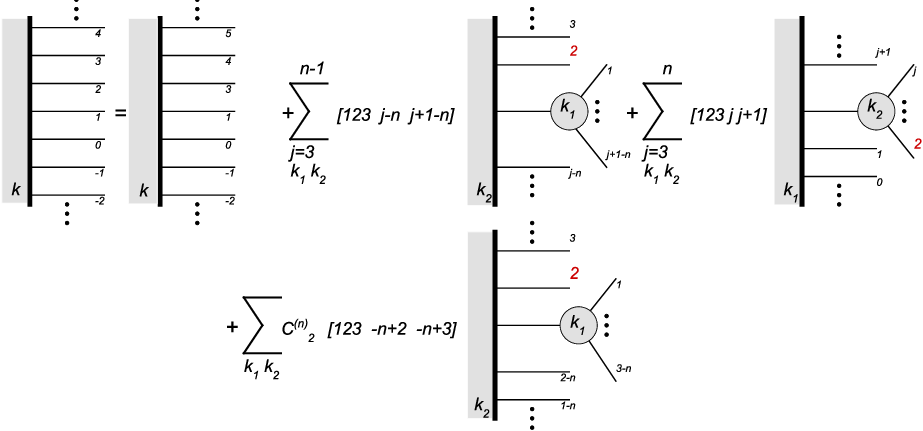}
 \end{center}\vspace{-0.2cm}
 \caption{Schematic representation of BCFW recursion for the
 $\mbox{N}^k\mbox{MHV}$ form factor $Z_n^{(0)(k)}$ at the tree level.
 The vertical bold black line corresponds to the form factor.
 The grey blob corresponds to the amplitude.
 $\mbox{MHV}$ form factors and amplitudes are factor out.
 In the last term the only non zero contribution is $k_2=0$, $k_1=k-1$.}\label{all_tree_FF}
 \end{figure}

Using the momentum supertwistors one can also easily write the
recursion relations for
$\mbox{N}^k\mbox{MHV}$ form factor $Z_n^{tree(k)}$ at tree level in full
analogy with
the amplitude case. Performing the following shift of
momentum supertwistor \cite{Henrietta_Amplitudes,Arkani_Hamed_Integrahd}
\begin{eqnarray}
\hat{\mathcal{Z}}_2=\mathcal{Z}_2+w\mathcal{Z}_3,
\end{eqnarray}
which is equivalent to the $[1,2\rangle$ shift in the momentum superspace
and considering integral:
\begin{eqnarray}
\oint \frac{d w}{w}\hat{Z}_n^{(k)}(w)=0,
\end{eqnarray}
one can obtain the following
recursion relations
($Z_n^{tree(0)}\equiv Z_n^{MHV}$,$A_n^{tree(0)}\equiv A_n^{MHV}$):
\begin{eqnarray}
&&\frac{Z_n^{(k)}(...,\mathcal{Z}_{-n+2},\mathcal{Z}_{-n+3},...,\mathcal{Z}_{1},\mathcal{Z}_{2},\mathcal{Z}_{3},
...,\mathcal{Z}_{n},\mathcal{Z}_{n+1},...)}{Z_n^{(0)}}=\nonumber\\
&=&\frac{Z_{n-1}^{(k)}}{Z_{n-1}^{(0)}}(...,\mathcal{Z}_{1-n},...,\mathcal{Z}_{1},\mathcal{Z}_{3},\mathcal{Z}_{4},...,\mathcal{Z}_{1+n},...)
\nonumber\\
&+&\sum_{j=3}^n[1,2,3,j,j+1]\times\frac{A_{n_1}^{(k_1)}}{A_{n_1}^{(0)}}
\left(\mathcal{Z}_I,\hat{\mathcal{Z}}_{2},...,\mathcal{Z}_{j}\right)\times
\frac{Z_{n_2}^{(k_2)}}{Z_{n_2}^{(0)}}
\left(...,\mathcal{Z}_{0},\mathcal{Z}_{1},\mathcal{Z}_I,\mathcal{Z}_{j+1},...\right)
\nonumber\\
&+&\sum_{j=3}^{n-1}[1,2,3,j-n,j+1-n]\times\frac{Z_{n_1}^{(k_1)}}{Z_{n_1}^{(0)}}
\left(...,\mathcal{Z}_{j-n},\mathcal{Z}_I,\hat{\mathcal{Z}}_{2},\mathcal{Z}_{3},...\right)
\times\frac{A_{n_2}^{(k_2)}}{A_{n_2}^{(0)}}
\left(\mathcal{Z}_I,\mathcal{Z}_{1},...,\mathcal{Z}_{j+1-n}\right)
\nonumber\\
&+&c^{(n)}_2[1,2,3,2-n,3-n]\times
\frac{Z_{2}^{(k_1)}}{Z_{2}^{(0)}}
\left(...,\mathcal{Z}_{2-n},\mathcal{Z}_I,\hat{\mathcal{Z}}_{2},\mathcal{Z}_{3},...\right)
\times
\frac{A_{n}^{(k_2)}}{A_{n}^{(0)}}\left(\mathcal{Z}_I,\mathcal{Z}_{1},...,\mathcal{Z}_{-n+3}\right).
\nonumber\\
\end{eqnarray}
with\footnote{$(jj+1)\bigcap(klm)=\mathcal{Z}_j\langle j+1klm\rangle
+\mathcal{Z}_{j+1}\langle jklm\rangle$}
\begin{eqnarray}
\mathcal{Z}_I=(jj+1)\bigcap(123)~\mbox{and}~\hat{\mathcal{Z}}_2=(12)\bigcap(0jj+1),
\end{eqnarray}
\begin{eqnarray}
n_1+n_2-2=n,~k_1+k_2+1=k.
\end{eqnarray}

These relations have a curious property that they represent the ratio
of $Z_n^{(k)}$ form factor and the MHV form factor in terms of
polynomials of the $[a,b,c,d,e]$ "brackets" multiplied by the coefficients
$c^{(k)}_p$ which are ratios of the $\langle a,b,c,d\rangle$ dual conformal
invariants. The $[a,b,c,d,e]$ bracket in the case of amplitudes is the dual
superconformal invariant.
In the case of the form factors we impose the $\gamma^{+}=0$ condition which
will likely brake some of the dual superconformal symmetries, but leave
ordinary dual conformal symmetry intact (?).
So the $[a,b,c,d,e]$ bracket in the case
of the form factors is the dual conformal invariant and
so is $c^{(k)}_p[a,b,c,d,e]$. The only inconsistency
which one can encounter is the behaviour of $c^{(k)}_p$
with respect to little group scaling \cite{Henrietta_Amplitudes}.
However, it is easy to see
that for the $n$ particle case if $Z_i$ and $Z_{i+nk}$, $k\in\mathbb{N}$
scaled the same way, which is expected, then $c^{(k)}_p$
is invariant with respect to little group scaling.
One may think that the ratio
of the $\mbox{N}^k\mbox{MHV}_n$ form factor and the $\mbox{MHV}_n$ form
factor at tree level is dual conformal invariant! It is
immediately tempting to speculate about the situation
at the loop level. At
one loop explicit answers are available for
$\mbox{N}\mbox{MHV}_{3,4}$. One may think that contributions from
$3m$ triangles will be an obstacle \cite{BORK_NMHV_FF}, it is unclear
at the first
glance how such contributions may cancel each other. This situation
as well as the symmetry properties of tree level form factors
require more detailed studies.

\section{Spurious poles cancellation, BCFW Vs all-line shift and polytopes}\label{p5}
So far we have formulated how to treat the form factors
in the momentum twistor space,
obtained BCFW recursion for the $\mbox{N}^k\mbox{MHV}_n$ form factor
in the momentum supertwistors representation,
and very briefly discussed their
possible symmetry properties.
The questions regarding BCFW and all-line shift (CSW)
equivalence and spurious poles cancellation remained unanswered.
However now we have all appropriate tools to address them.

At first, let us try to see that BCFW and all-line shift
(CSW) recursion are equivalent, at least in NMHV sector.
Here we aim at the concrete examples rather then general proofs,
and will consider mostly $n=3,4$ $\mbox{N}\mbox{MHV}$ cases.

Let us rewrite all-line shift (CSW) results
for the $\mbox{N}\mbox{MHV}$ sector in the momentum
supertwistors. One can obtain \cite{HarmonyofFF_Brandhuber}:
\begin{eqnarray}
Z^{NMHV}_n/Z^{MHV}_n=\sum_{i=1}^n\sum_{j=i+2}^{i+n-1}[*,i,i+1,j,j+1].
\end{eqnarray}
Here $\mathcal{Z}^*$ is an arbitrary supertwistor with components
$\lambda^*=\chi^*=0$. One can choose
$\mu^{*}=\tilde{\lambda}^{*}$. The $\gamma^+ \to 0$ condition is implemented.

One can also think that $Z^*$ is an obtained from a twistor
with arbitrary components by contraction with the so called infinity
twistor $I^{AB}$ \cite{Masson_Skiner_Grassmaians_Twistors}. The presence of
the infinity twistor explicitly brakes dual conformal invariance of each
$[*,a,b,c,d]$ term in the
all-line shift (CSW) representation of the amplitude or the form factor.
In the case of
amplitudes, dual conformal invariance is restored in the whole sum of the
$[*,a,b,c,d]$
terms. We expect a similar situation in the case of the form factors.

Note that the form of the all-line shift (CSW) representation discussed here
is not unique, due to the periodical nature of the contour.
One can start the first sum ("fix the gauge")
$\sum_{i=1}^n$ from an arbitrary point on the contour,
for example, from
$i=-1$: $\sum_{i=-1}^{n-2}\sum_{j=i+2}^{i+n-1}$; this will lead to the same
formula if one will
return from momentum twistors to momentum superspace variables,
as was explained earlier.
It is convenient to "fix the gauge" this way in our case,
i.e., start summation from the point $i=-1$:
 \begin{eqnarray}
Z^{NMHV}_n/Z^{MHV}_n=\sum_{i=-1}^{n-2}\sum_{j=i+2}^{i+n-1}[*,i,i+1,j,j+1].
\end{eqnarray}
 Equivalently, we can shift ("fix another gauge") our BCFW results by
 appropriate amount of
 periods, but we will not do so.
 Then for $n=3$ and $n=4$ one can write:
\begin{eqnarray}
Z^{NMHV}_3/Z^{MHV}_3=[*,-1,0,1,2]+[*,0,1,2,3]+[*,1,2,3,4],
\end{eqnarray}
and
\begin{eqnarray}
Z^{NMHV}_4/Z^{MHV}_4&=&([*,-1,0,1,2]+[*,-1,0,2,3])+([*,0,1,2,3]+[*,0,1,3,4])+
\nonumber\\
&+&([*,1,2,3,4]+[*,1,2,4,5])+([*,2,3,4,5]+[*,2,3,5,6]).\nonumber\\
\end{eqnarray}

Our next step is to show the sketch of the proof that the following
equality holds :
\begin{eqnarray}
c^{(n)}_i[1,i,i+1,i-n,i+1-n]&=&[*,1,i,i+1,1+n]+[*,1,i,i-n,i+1-n]\nonumber\\
&+&[*,1,i,i+1,i+1-n],
\end{eqnarray}
$\gamma^+ \to 0$ condition is implemented, $\chi^*=0$ and $Z^*$ is the result
of projection by means of the infinity twistor $I^{AB}$.
One can think about it as some kind of partial fractions decomposition.
Let us proceed by iterations. For $n=3$ one can verify that this equality
holds by explicit comparison of the coefficients before Grassmann monomials.
For example,
\begin{eqnarray}\label{3term}
c^{(3)}_2[-1,0,1,2,3]&=&[*,-1,0,1,2]+[*,0,1,2,3]+[*,1,2,3,4],\nonumber\\
c^{(3)}_2&=&\frac{\langle-1,1,2,3\rangle\langle-1,0,1,3\rangle}{\langle-1,0,2,3\rangle\langle1,2,3,4\rangle}.
\end{eqnarray}
Note that LHS of the equality has poles $\langle-1,0,1,2\rangle$,
$\langle0,1,2,3\rangle$,
and $\langle1,2,3,4\rangle$. The pole $\langle-1,0,1,3\rangle \sim q^2$
as was explained earlier is absent.
In RHS we separated these poles by introducing the $\mathcal{Z}^*$ axillary
supertwistor.
In fact for $n=3$ this equality is just a statement
that BCFW and  all-line shift (CSW)
gives the same result:
\begin{eqnarray}
Z^{NMHV}_{3,BCFW}=Z^{NMHV}_{3,CSW}.
\end{eqnarray}
One can also check that the dependence
on the axillary twistor is canceled in all coefficients.
Then we can substitute in the BCFW recursion for $n=4$ in the term
\begin{eqnarray}
\left(Z_{3}^{(0)NMHV}\otimes A_3^{(0)\overline{MHV}}\right)
\end{eqnarray}
$Z^{NMHV}_{3}$ in the form $Z^{NMHV}_{3,BCFW}$ or $Z^{NMHV}_{3,CSW}$.
Comparing two results and considering all possible
$[i,j\rangle$ shifts we can prove the identity (\ref{3term}) for $n=4$.
Then we can substitute in BCFW recursion for $n=5$ the results obtained
for $n=4$, etc.

Now one can see that substituting in BCFW formula identity (\ref{3term})
containing the axillary supertwistor $\mathcal{Z}^*$
and using the 6 term identity \cite{Hoges_Polytopes,Masson_Skiner_Grassmaians_Twistors}
for the set of twistors
\begin{eqnarray}
\mathcal{Z}^*,\mathcal{Z}_1,\mathcal{Z}_a,
\mathcal{Z}_b,\mathcal{Z}_c,\mathcal{Z}_d
\end{eqnarray}
for all other $[1,a,b,c,d]$ invariants:
\begin{eqnarray}
[1,a,b,c,d]=[*,a,b,c,d]-[*,1,b,c,d]+[*,1,a,c,d]-[*,1,a,b,d]+[*,1,a,b,c],
\nonumber\\
\end{eqnarray}
the all-line shift (CSW) formula is reproduced.
Let us illustrate this by the $n=4$ example. Substituting
\begin{eqnarray}
[1,2,3,4,5]=[*,2,3,4,5]-[*,1,3,4,5]+[*,1,2,4,5]-[*,1,2,3,5]+[*,1,2,3,4],
\nonumber\\
\end{eqnarray}
\begin{eqnarray}
[-1,0,1,2,3]&=&[*,0,1,2,3]-[*,-1,1,2,3]+[*,-1,0,2,3]-[*,-1,0,1,3]
\nonumber\\&+&[*,-1,0,1,2],
\nonumber\\
\end{eqnarray}
\begin{eqnarray}
(\mathbb{S}c^{(3)}_2)[-1,0,1,3,4]=[*,-1,0,1,3]+[*,0,1,3,4]+[*,1,3,4,5],
\end{eqnarray}
\begin{eqnarray}
c^{(4)}_2[1,2,3,-2,-1]=[*,-2,-1,1,2]+[*,-1,1,2,3]+[*,1,2,3,5],
\end{eqnarray}
in the BCFW result one obtains ($[*,-2,-1,1,2]=[*,2,3,5,6]$ for $n=4$)
\begin{eqnarray}
Z^{NMHV}_4/Z^{MHV}_4&=&[*,2,3,4,5]+[*,1,2,4,5]+[*,1,2,3,4]+[*,0,1,2,3]+[*,-1,0,2,3]
\nonumber\\
&+&[*,-1,0,1,2]+[*,0,1,3,4]+[*,2,3,5,6].
\end{eqnarray}
which is  the all-line shift (CSW) formula.

So far we argued how to transform the BCFW representation of
$\mbox{NMHV}$ form factors into the all-line shift (CSW) one. But what
about cancelation of spurious poles ? Let us start with the $n=4$ point
example, as an illustration, how spurious pole cancels. As it was
explained earlier, one of the spurious poles $\langle1|q|2]$ should
be canceled between the terms
\begin{eqnarray}
\tilde{R}_{122}^{(1)}=c^{(4)}_2[1,2,3,-2,-1] ~\mbox{and}~
R_{142}^{(2)}=[1,2,3,4,5].
\end{eqnarray}
Let us consider a component expression proportional to
$\chi_5^{-}\chi_5^{-}\chi_2^+\chi_3^+$. Note also that ($\chi_2^{+}=\chi_{-2}^{+}$, $\chi_3^{+}=\chi_{-1}^{+}$ because
$\gamma^+=0$) the coefficient of $\chi_5^{-2}\chi_2^+\chi_3^+$ should be equivalent to coefficient before $\chi_1^{-2}\chi_{-2}^+\chi_{-1}^+$ due to the periodical nature of the contour.
Extracting the corresponding components we see that (here we drop $\mp$ subscript):
\begin{eqnarray}
\left.[1,2,3,4,5]\right|_{\chi_5^2\chi_2\chi_3}=\frac{\langle
1,2,3,4\rangle}{\langle3,4,5,2\rangle\langle5,1,2,3\rangle},
\end{eqnarray}
and
\begin{eqnarray}
\left.c^{(4)}_2[1,2,3,-2,-1]\right|_{\chi_1^2\chi_{-2}\chi_{-1}}
=\left.\left(\mathbb{P}^4c^{(4)}_2\right)[2,3,5,6,7]\right|_{\chi_5^2\chi_2\chi_3}=
\frac{\langle2,5,6,7\rangle}{\langle1,2,3,5\rangle\langle2,3,5,6\rangle}.\nonumber\\
\end{eqnarray}
So for the form factor we have
\begin{eqnarray}
\left.Z^{NMHV}_4/Z^{MHV}_4\right|_{\chi_1^2\chi_{-2}\chi_{-1}}=
\frac{1}{\langle1,2,3,5\rangle}
\left(\frac{\langle1,2,3,4\rangle}{\langle2,3,4,5\rangle}
+\frac{\langle2,5,6,7\rangle}{\langle2,3,5,6\rangle}\right).
\end{eqnarray}
$\langle1,2,3,5\rangle\sim\langle1|q|2]$, and we see that if the expression
in the brackets vanishes as $\langle1,2,3,5\rangle \rightarrow 0$, then
$\langle1,2,3,5\rangle$ pole is canceled exactly as in the \cite{Hoges_Polytopes}
example. Using identity for 6 twistors
$Z_{1},\ldots,Z_{5},Z_{X}$:
\begin{eqnarray}
\langle2,3,1,4\rangle\langle2,3,5,X\rangle
+\langle2,3,1,5\rangle\langle2,3,4,X\rangle
+\langle2,3,1,X\rangle\langle2,3,4,5\rangle=0
\end{eqnarray}
One can see that as $\langle1,2,3,5\rangle \rightarrow 0$
\begin{eqnarray}
\frac{\langle2,3,1,4\rangle}{\langle2,3,4,5\rangle}=
\frac{\langle2,3,1,X\rangle}{\langle2,3,5,X\rangle}.
\end{eqnarray}
This identity is valid for arbitrary 6 twistors, so we can choose
$Z_{X}=Z_{6}$. 
Using identity (\ref{Relation_<abcd>_2}) which in our case gives us $\langle5,6,7,2\rangle=\langle1,2,3,6\rangle$ one
can see that indeed as $\langle1,2,3,5\rangle \rightarrow 0$ expression
in brackets cancels. This is a good sign, but one would like to have
more general statement regarding the spurious pole cancelation.

Transforming the BCFW representation into CSW we recast all BCFW
spurious poles into poles containing the $\mathcal{Z}^*$ twistor:
$\langle*,a,b,c\rangle$. We also get rid of the terms with the coefficients
$c^{(n)}_i$, so our answer is represented only as the sum
of $[*,a,b,c,d]$ invariants.

In the amplitude case, one can use the
geometrical interpretation of the amplitude as the volume of a polytope
in $\mathbb{C}\mathbb{P}^4$ to show that all poles
of the form $\langle*,a,b,c\rangle$
cancel \cite{Arcani_Hamed_Polytopes}.
The $[a,b,c,d,e]$ invariant is interpreted as the volume
of 4-simplex in $\mathbb{C}\mathbb{P}^4$
\cite{Henrietta_Amplitudes,Arcani_Hamed_Polytopes}.
The $\mbox{NMHV}$ amplitude is the sum of volumes of such 4-simplixes,
and hence can be interpreted as the volume of the polytope.
The 4-simplixes in BCFW or all-line shift (CSW) recursion represents
particular triangulation of this polytope.
The poles in $[a,b,c,d,e]$ are "brackets" of the form $\langle a,b,c,d\rangle$
which correspond to the vertexes of the 4-simplex in the geometrical picture.
Cancellation
of spurious poles can be seen in this picture as
"cancellation" of the contribution of
the corresponding vertices: 4-simplexes are combined into
a polytope (amplitude) in such a way that the resulting
polytope (amplitude) will have only such vertexes that correspond
to the physical poles.

Our aim now is to show that the same ideas about
the spurious pole cancellation can be applied to the form factors as well,
with some minor but curious changes.

First of all, let us explain
how one can rewrite $[a,b,c,d,e]$ invariants as volumes
of the $\mathbb{C}\mathbb{P}^4$ simplexes in the case when we
are dealing with the harmonic superspace.
We introduce new fermionic variables $X^{+a}$ and $X^{-a'}$
\begin{equation}
X^{+a}\chi^{-}_{ai}=\psi_i^{(-)},~X^{-a'}\chi^{+}_{a'i}=\psi^{(+)}_i
\end{equation}
such that $\psi^{(-)}_i=\psi^{(+)}_i$. Here the $(\pm)$
subscript stands to distinguish dependence
of $\psi$ and other objects on $\chi^{-}$  or $\chi^{+}$.
Then we can introduce 5 component objects which
we will treat as the set of homogeneous coordinates on $\mathbb{C}\mathbb{P}^4$
\begin{eqnarray}
\mathbf{Z}_i^{(\pm)}=(Z_i,\psi^{(\pm)}_i) \mbox{ -- 5 comopnent object},
\end{eqnarray}
and
\begin{eqnarray}
\mathbf{Z}_0=(0,0,0,0,1),
\end{eqnarray}
such that
\begin{eqnarray}
\hat{\delta}^{\pm2}(\chi^{\pm}_a\langle b,c,d,e\rangle + \mbox{cycl.})&=&\frac{1}{2!}
\int d^{\pm2}X ~\langle a,b,c,d,e \rangle^{2(\pm)},
\nonumber\\ \langle a,b,c,d\rangle&=&\langle\mathbf{0},a,b,c,d\rangle \equiv \langle\mathbf{0},a,b,c,d\rangle^{(\pm)},
\end{eqnarray}
where
\begin{eqnarray}
\langle a,b,c,d,e \rangle^{(\pm)} = \epsilon_{q_1q_2q_3q_4q_5}\mathbf{Z}^{(\pm) q_1}_{a}\mathbf{Z}^{(\pm) q_2}_{b}
\mathbf{Z}^{(\pm) q_3}_{c}\mathbf{Z}^{(\pm) q_4}_{d}\mathbf{Z}^{(\pm) q_5}_{e},
\end{eqnarray}
\begin{eqnarray}
\langle \textbf{0},b,c,d,e \rangle^{(\pm)} = \epsilon_{q_1q_2q_3q_4q_5}\mathbf{Z}^{q_1}_{0}\mathbf{Z}^{(\pm)q_2}_{b}
\mathbf{Z}^{(\pm)q_3}_{c}\mathbf{Z}^{(\pm)q_4}_{d}\mathbf{Z}^{(\pm)q_5}_{e}.
\end{eqnarray}
Since in the case of amplitudes $\psi^{(-)}_i=\psi^{(+)}_i$, we have
$\langle a,b,c,d,e \rangle^{2(-)}=\langle a,b,c,d,e \rangle^{2(+)}$, so
\begin{eqnarray}
\hat{\delta}^{4}(\chi_a\langle b,c,d,e\rangle + \mbox{cycl.})=\frac{4!}{2!2!}
\int d^{-2}X d^{+2}X ~\frac{1}{4!}\langle a,b,c,d,e \rangle^{4},
\end{eqnarray}
and we can rewrite
$[a,b,c,d,e]$ in the following way
($\int_{X}\equiv 4!/2!2!\int d^{-2}X d^{+2}X$):
\begin{eqnarray}
[a,b,c,d,e]\equiv\int_{X}
\frac{1}{4!}\frac{\langle a,b,c,d,e\rangle^4}{\langle\textbf{0},a,b,c,d\rangle
\langle\textbf{0},b,c,d,e\rangle\langle\textbf{0},c,d,e,a\rangle\langle\textbf{0},d,e,a,b,\rangle\langle\textbf{0},e,a,b,c\rangle}.
\nonumber\\
\end{eqnarray}
Comparing this with the formula for the volume of the 4-simplex
in $\mathbb{C}\mathbb{P}^4$
\begin{eqnarray}
Vol_4[a,b,c,d,e]=\frac{1}{4!}\frac{\langle a,b,c,d,e\rangle^4}{\langle\textbf{0},a,b,c,d\rangle
\langle\textbf{0},b,c,d,e\rangle\langle\textbf{0}c,d,e,a\rangle\langle\textbf{0},d,e,a,b\rangle\langle\textbf{0},e,a,b,c\rangle},
\end{eqnarray}
we see that
\begin{eqnarray}
[a,b,c,d,e]=\int_{X} Vol_4[a,b,c,d,e].
\end{eqnarray}
One can see that the NMHV amplitude is given by the sum of $Vol_4$.
Let us also write for comparison the general formula for volume of the simplex in $\mathbb{C}\mathbb{P}^n$
\begin{eqnarray}
Vol_n(a_1,\ldots,a_{n+1})=\frac{1}{n!}\frac{\langle a_1,\ldots,a_{n+1}\rangle^n}{\langle\textbf{0},a_1,\ldots,a_n\rangle
\ldots\langle\textbf{0},a_{n+1},a_1,\ldots,a_{n-1}\rangle}.
\end{eqnarray}

To get some geometrical intuition how this volume formula works, consider $\mathbb{C}\mathbb{P}^2$ case \cite{Henrietta_Amplitudes}:
\begin{eqnarray}
Vol_2[a,b,c]=\frac{1}{2!}\frac{\langle a,b,c\rangle^2}{\langle\textbf{0},a,b\rangle
\langle\textbf{0},b,c\rangle\langle\textbf{0},c,a\rangle}.
\end{eqnarray}
The 3 component objects $Z_a^{I},Z_b^{I},Z_c^{I}$, $I=1,\ldots,3$,
which are homogeneous coordinates on $\mathbb{C}\mathbb{P}^2$ define 3 lines in the dual $\mathbb{C}\mathbb{P}^2$
space, with the coordinates $W_I$, via the conditions
\footnote{We are considering the projective geometry, so if one will consider
$W$ as points in the 3 dimensional affine spaces $\bold{W}$, condition $(ZW)=0$,
for fixed $Z$ defines a plane in $\bold{W}$.
Intersection of this plane with the plane
defined by $Z_{\textbf{0}}$ gives us line,
which we are talking about.} $(Z_a W) \equiv Z_a^{I}W_I=0$.
In $\mathbb{C}\mathbb{P}^n$ $Z$ will define the $n-1$ subspace.
In the $\mathbb{C}\mathbb{P}^2$ case these lines,
defined by $Z_a^{I},Z_b^{I},Z_c^{I}$ intersect at the points
\begin{eqnarray}
W_{1I}&=&W_{(ab)I}=\epsilon_{IJK}Z_a^{J}Z_b^{K},\nonumber\\
W_{2I}&=&W_{(bc)I}=\epsilon_{IJK}Z_b^{J}Z_c^{K},\nonumber\\
W_{3I}&=&W_{(ca)I}=\epsilon_{IJK}Z_c^{J}Z_a^{K}.
\end{eqnarray}
These points are projected on a plane defined by $Z_{\textbf{0}}$, and one can think of
them as vertices of 2d triangle (two dimensional simplex), with the edges defined by $Z_a^{I},Z_b^{I},Z_c^{I}$;
$Vol_2[a,b,c]$ is the projectively defined
(it is invariant under rescalings of $Z^I \to \lambda Z^I$
or $W_I \to \lambda W_I$, while $Z_{\textbf{0}}$ is always fixed,
$\lambda$ is some number) area of this triangle.
The vertexes of this triangle are in one to one
correspondence with
$\langle\textbf{0},a,b\rangle$, etc.
"scalar products". In terms of $W$'s
$Vol_2[a,b,c]$ is given by
($(Z_{\textbf{0}}W_1)=\langle\textbf{0},a,b\rangle$, etc.)
\begin{eqnarray}
Vol_2[a,b,c]=\frac{1}{2!}\frac{\langle W_1,W_2,W_3\rangle}{(Z_{\textbf{0}}W_1)(Z_{\textbf{0}}W_2)(Z_{\textbf{0}}W_3)}.
\end{eqnarray}
Using projective invariance $W_I \to \lambda W_I$ one can
always choose $W_1,W_2,W_3$ in the form
$W_1=(x_1,y_1,1)$, $W_2=(x_2,y_2,1)$, $W_3=(x_3,y_3,1)$.
$x_i,y_i$ are then the coordinates of the
vertices of $(a,b,c)$ triangle in the plane defined by $Z_{\textbf{0}}$.
\begin{figure}[t]
 \begin{center}
  \epsfxsize=8cm
 \epsffile{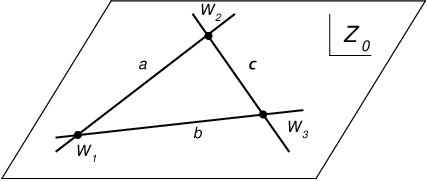}
 \end{center}\vspace{-0.2cm}
 \caption{$\mathbb{C}\mathbb{P}^2$ Simplex defined by
 $Z_a,Z_b,Z_c$.}\label{CP2_simplex}
 \end{figure}

The situation when one of the brackets in the denominator
(for example $\langle\textbf{0},a,b\rangle=0$) is equal to
0 corresponds
in general to the case when $W_{1}$ point moves to infinity so that
$Vol_2[a,b,c]$ becomes singular (infinite).

In the $\mathbb{C}\mathbb{P}^4$ case, we are really interested in,
the $Z$ twistors define three dimensional subspaces
in dual the $\mathbb{C}\mathbb{P}^4$ space.
Intersections of these three dimensional subspaces define vertices
of the four dimensional simplex. The vertexes of this simplex are
in one-to-one correspondence with
$\langle\textbf{0},a,b,c,d\rangle=\langle a,b,c,d\rangle$ poles.

\begin{figure}[t]
 \begin{center}
  \epsfxsize=14cm
 \epsffile{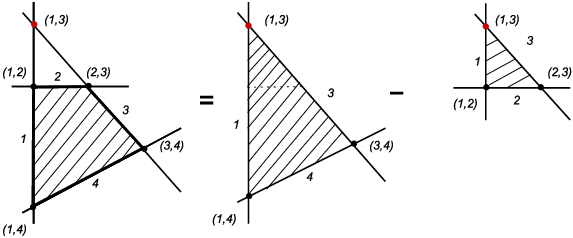}
 \end{center}\vspace{-0.2cm}
 \caption{Cancellation of $(1,3)$ pole in $Vol_2[1,4,3]-Vol_2[1,2,3]$.}
 \label{pole_cancellation_fig}
 \end{figure}

To see how one can observe cancellation of poles (vertices) in this
geometrical picture, let us return to the $\mathbb{C}\mathbb{P}^2$ example
\cite{Henrietta_Amplitudes}.
Consider two triangles defined by $Z_1,Z_2,Z_3$ and $Z_1,Z_3,Z_4$.
In the difference $Vol_2[1,2,3]-Vol_2[1,4,3]$ ($Vol_2[1,4,3]=-Vol_2[1,3,4]$)
the contribution of the
$\langle\textbf{0},1,3\rangle$ vertex will drop out, so the difference is
regular in the $\langle\textbf{0},1,3\rangle \to 0$ limit.
See fig.\ref{pole_cancellation_fig}.

To see this cancellation in a more algebraic way, without drawing pictures,
which is very convenient when we are dealing with four dimensional
volumes, let us introduce a boundary operator $\partial$ for the simplex
in $\mathbb{C}\mathbb{P}^n$ which gives the volume of the
boundary of this simplex (i.e. combination of volumes of the simplexes in
$\mathbb{C}\mathbb{P}^{n-1}$) \cite{Henrietta_Amplitudes}:
\begin{eqnarray}
\partial Vol_n[1,2,3,...,n]=
\sum_{i=1}^n(-1)^{i+1}Vol_{n-1}[1,2,...,i-1,i+1,...,n]|^{Z_i}.
\end{eqnarray}
One can verify that as expected $\partial^2=0$,
$Vol_{n-1}[1,2,...,i-1,i+1,...,n]|^{Z_i}$
is defined as the projection of the $(1,2,...,i-1,i+1,...,n)$
lines into the $n-1$ dimensional
subspace defined by $Z_i$. Returning to the $\mathbb{C}\mathbb{P}^2$
case one can see that
\begin{eqnarray}
\partial Vol_2[1,2,3]&=&Vol_1[2,3]|^{Z_1}-Vol_1[1,3]|^{Z_2}+Vol_1[1,2]|^{Z_3},\nonumber\\
\partial Vol_2[1,3,4]&=&Vol_1[4,3]|^{Z_1}-Vol_1[1,4]|^{Z_3}+Vol_1[1,3]|^{Z_4}.
\end{eqnarray}
The boundaries (line segments) of the triangles
$Vol_1[1,3]|^{Z_2}$ and $Vol_1[1,3]|^{Z_4}$ corresponding to the
$\langle \textbf{0}13\rangle$ vertex (pole) encounters with
the opposite sign.
This
corresponds to the situation when such vertex is absent in the
final polytope (sum of simplexes).
The same will be true in the general case of the sum of the simplexes in
$\mathbb{C}\mathbb{P}^{n-1}$.

In summary \cite{Henrietta_Amplitudes,Arcani_Hamed_Polytopes}
one can say that to figure out which vertices (poles) will be present
in the polytope combined from the set of simplexes, one has to act with
the boundary operator $\partial$ on each simplex
and "cancel" all vertices with the opposite sign ignoring the $|^{Z_i}$ subscript.
Hereafter we will drop the
$|^{Z_i}$ subscript.

As an example, one can check that in the case of the
all-line shift (CSW) representation of the $n=5$
NMHV amplitude in the result of the action of the boundary operator on
the individual simplexes,
all the poles (vertices) of the form $\langle\textbf{0},*,a,b,c\rangle$
are "canceled" and only physical poles of the form
$\langle\textbf{0},a,b,c,d\rangle$ remain. This also reflects
the fact that the result should be independent
of the explicit choice of $\mu^{*}$ in $\mathcal{Z}^*$.
In fact in the case of
amplitudes one can see that the result is independent of the
choice of all components in
$\mathcal{Z}^*$ recasting the all-line shift (CSW) representation
into the BCFW one by using the 6 term identity.

Now let us return to the form factors. Due to the presence of
$\gamma^{+}=0$ condition
on the periodical contour (fermionic part $\chi^+_i$ of the contour is closed)
$\psi^{(-)}_i \neq \psi^{(+)}_i$. So in the case of the form factors one can
write
\begin{eqnarray}
[a,b,c,d,e]&\equiv&\int_{X} \frac{1}{4!}\frac{\langle a,b,c,d,e\rangle^{(-)2}}
{(\langle\textbf{0},a,b,c,d\rangle
\langle\textbf{0},b,c,d,e\rangle\langle\textbf{0},c,d,e,a\rangle\langle\textbf{0},d,e,a,b\rangle\langle\textbf{0},e,a,b,c\rangle)^{1/2}}
\nonumber\\
&\times&\frac{\langle a,b,c,d,e\rangle^{(+)2}}{(\langle\textbf{0},a,b,c,d\rangle
\langle\textbf{0},b,c,d,e\rangle\langle\textbf{0},c,d,e,a\rangle\langle\textbf{0},d,e,a,b\rangle\langle\textbf{0},e,a,b,c\rangle)^{1/2}}
\nonumber\\&=&
\int_{X}\left(Vol_4[a,b,c,d,e]^{(-)}\right)^{1/2}\left(Vol_4[a,b,c,d,e]^{(+)}\right)^{1/2}.
\end{eqnarray}
The only difference in $(Vol_4[a,b,c,d,e]^{(-)})^{1/2}$
and $(Vol_4[a,b,c,d,e]^{(+)})^{1/2}$
is the fermionic components $\chi^+_i$ and $\chi^-_i$.
As it is not convenient to work
with square roots of volumes one can consider axillary objects where
$\gamma^-$ and $\gamma^+$
(hence $\chi^+_i$ and $\chi^-_i$) enter on an equal footing
and the limit
$\gamma^+ \to 0$ is taken only in the final result.
As it was explained before, this limit is not singular. If some poles
cancel in the sum of $[a,b,c,d,e]$ before the $\gamma^+ \to 0$
limit they also should
cancel after this limit is taken;
$[a,b,c,d,e]$ are ratio of polynomials.
So if in the sum
of such ratio
of polynomials some poles of individual terms cancel,
taking one coefficient to 0 in the numerators
of such polynomials should not affect pole cancellation.
From this point of view the $\mbox{NMHV}$ form factor is not exactly
the $\mathbb{C}\mathbb{P}^4$ polytope but rather its special
limit ($\gamma^+ \to 0$).

Now consider the three point NMHV form factor
(here we choose the contour periods as in \cite{HarmonyofFF_Brandhuber})
\begin{eqnarray}
Z_3^{NMHV}/Z_3^{MHV}=[*,0,1,2,3]+[*,1,2,3,4]+[*,2,3,4,5].
\end{eqnarray}
Considering $\gamma^+\neq 0$ let us apply the boundary operator to the
individual terms:
\begin{eqnarray}
\partial Vol_4[*,0,1,2,3]&=& Vol_3[0,1,2,3]-Vol_3[*,1,2,3]+
(Vol_3[*,0,2,3]-Vol_3[*,0,1,3])\nonumber\\&+&Vol_3[*,0,1,2],\nonumber\\
\partial Vol_4[*,1,2,3,4]&=& Vol_3[1,2,3,4]-Vol_3[*,2,3,4]+(Vol_3[*,1,3,4]-Vol_3[*,1,2,4])
\nonumber\\&+&Vol_3[*,1,2,3],\nonumber\\
\partial Vol_4[*,2,3,4,5]&=& Vol_3[2,3,4,5]-Vol_3[*,3,4,5]+(Vol_3[*,2,4,5]-Vol_3[*,2,3,5])
\nonumber\\&+&Vol_3[*,2,3,4].\nonumber\\
\end{eqnarray}
We see that the poles corresponding to
$Vol_3[*,0,2,3]$, $Vol_3[*,0,1,3]$, $Vol_3[*,1,3,4]$,
$Vol_3[*,1,2,4]$, $Vol_3[*,2,4,5]$, $Vol_3[*,2,3,5]$
are not canceled in such axillary object.
All other poles are canceled
(Note that $Vol_3[*,3,4,5]$ and $Vol_3[*,0,1,2]$ correspond to the same
pole in the $n=3$ case).
We also see that
\begin{eqnarray}
Vol_3[*,1,3,4]-Vol_3[*,1,2,4]=\mathbb{P}(Vol_3[*,0,2,3]-Vol_3[*,0,1,3]),
\end{eqnarray}
and
\begin{eqnarray}
Vol_3[*,2,4,5]-Vol_3[*,2,3,4]=\mathbb{P}^2(Vol_3[*,0,2,3]-Vol_3[*,0,1,3]).
\end{eqnarray}
So if these poles are canceled in the first term, they will be canceled in other terms as well.

Now what will change if we take the $\gamma^+ \to 0$ limit ?
First of all let us note that the
$\langle\textbf{0},*,0,2,3 \rangle=\langle*,0,2,3 \rangle$
and $\langle\textbf{0},*,0,1,3 \rangle=\langle*,0,1,3 \rangle$ vertices in
fact correspond to the same pole $[*|q|3\rangle$:
\begin{eqnarray}
\langle*,0,2,3 \rangle&=&\langle23\rangle [*|x_{30}|3\rangle=\langle23\rangle [*|q|3\rangle,\nonumber\\
\langle*,0,1,3 \rangle&=&\langle31\rangle [*|x_{13}|3\rangle=\langle31\rangle [*|q|3\rangle.
\end{eqnarray}
The formulas
\begin{eqnarray}
\langle *,i-1,i,j\rangle=\langle i-1i\rangle [*|x_{ij}|j\rangle,~x_{ij}=\sum_{k=i}^{j-1}p_k,
\end{eqnarray}
\begin{eqnarray}
Z_{i}=(\lambda_{i},x_{i}\lambda_{i}),~
Z_{i+nk}=(\lambda_{i},x_{i+nk}\lambda_{i}),~i=1...n,~k\in\mathbb{N}.
\end{eqnarray}
were used.
Now consider the argument of the $\hat{\delta}^{+2}$ function
in $[*,0,1,2,3]$ in more detail. The argument
looks like (note that $\chi_*^+=0$)
\begin{eqnarray}
\chi_3^+\langle*,0,1,2\rangle+\chi_0^+\langle*,1,2,3\rangle+
\chi_1^+\langle*,0,2,3\rangle+\chi_2^+\langle*,0,1,3\rangle,
\end{eqnarray}
$\gamma^+=0$ corresponds to $\chi^+_0=\chi^+_3$, so we can write
the argument of the delta function as
\begin{eqnarray}
\chi_3^+(\langle*,0,1,2\rangle+\langle*,1,2,3\rangle)+
\chi_1^+\langle*,0,2,3\rangle+\chi_2^+\langle*,0,1,3\rangle.
\end{eqnarray}
For $\langle*,0,1,2\rangle$ and $\langle*,1,2,3\rangle$ one can get
\begin{eqnarray}
\langle*,0,1,2\rangle+\langle*,1,2,3\rangle=[*|q|3\rangle \langle12\rangle,
\end{eqnarray}
so $[*|q|3\rangle$ factors out from the delta function
and one can see that $[*|q|3\rangle^2 \sim \hat{\delta}^{+2}$.
The poles $\langle*,0,2,3 \rangle \sim [*|q|3\rangle$ and
$\langle*,0,1,3 \rangle \sim [*|q|3\rangle$
are exactly canceled! This is similar to the cancelation of
$q^2$ pole in $\tilde{R}^{(1)}_{rtt}$.
Note that such factorisation is possible only
in the $\gamma^+ \to 0$ limit. $\hat{\delta}^{-2}$
dose not factories in such a way.
From a geometric point of view this means that as
$[*|q|3\rangle \to 0$ $Vol_4[*,0,1,2,3]^{(-)}$
becomes singular, while $Vol_4[*,0,1,2,3]^{(+)} \to 0$
so that their product remains finite.
Such cancellation of the poles is the general pattern for all
$[*,a,b,c,d]$ coefficients with $a=i$
and $b=i\pm n$ for the $n$ point form factor.

For the general $n$ the situation is the same as in the $n=3$ example
and all poles
containing the $\mu^*$ dependence, except pairs of poles which come from the
$[*,a,b,c,d]$ coefficients with $a=i$ and $b=i\pm n$, are already cancel in
the axillary expression with $\gamma^+ \neq 0$.
The remaining pairs of poles cancel in $\gamma^+ \to 0$
limit. In the appendix one can find the details on the $n=4$ example.

Summing up, for $Z_n^{NMHV}/Z_n^{MHV}$ in the all-line shift (CSW) representation
the answer is free from poles containing the $\mathcal{Z}^*$ dependence which also
imply cancellation of spurious poles
in BCFW picture and independence of all-line shift (CSW) result
on the choice of $\mu^*$. This cancellation most easily can be seen
geometrically when we represent the
$[*,a,b,c,d]$ invariants as the volumes or the products of volumes of
the simplexes in $\mathbb{C}\mathbb{P}^4$.
This situation is similar to the amplitude case,
but there are some differences unique to the form factors due to
their special Grassmann structure. The form factor is not exactly the
$\mathbb{C}\mathbb{P}^4$ polytope but rather special
limit ($\gamma^+ \to 0$) of such polytope.

\section{Conclusion}\label{p6}
In this article, we considered different types of recursion relations for
the form factors of operators from the stress tensor supermultiplet in the
$\mathcal{N}=4$ SYM theory. We formulated the BCFW recursion relations in the
momentum twistor space for general helicity configuration and considered
the NMHV sector in more details. Using the momentum twistor space representation
we demonstrated the equivalence between the BCFW and all-line shift (CSW)
recursion relations
at least for the NMHV sector and used geometrical interpretation
of the NMHV form factors as the volumes of polytopes to show that the BCFW/all-line shift (CSW)
representations of the form factors are free from spurious poles.
The relation between the logarithmical derivative
of the form factor with respect to the coupling constant
and the amplitudes were also considered.
In addition,
we briefly discussed how the momentum twistor representation can be used
to clarify the relation between the IR pole coefficients at the one loop level. We hope
that similar ideas can be used beyond the NMHV sector.

The main conceptual result of this article is that the "on-shell structures
and ideas" such as the momentum twistor representation,
Yangian momentum twistor invariant function $[abcde]$ or the
polytope interpretation of the NMHV amplitudes still
play an essential role for partially off-shell objects such
as the form factors (or at least for the form factors of
operators from the stress tensor supermultiplet).
However, several important questions still remain unanswered.

It is well known that different BCFW shifts give representations
of the same amplitude, which looks different at the first glance.
For example, for the NMHV sector six point amplitude we have for
the $[1,2\rangle$ shift:
\begin{eqnarray}
 \frac{A_6^{NMHV}}{A_6^{MHV}}=
 [1,2,3,4,5]+[1,2,3,5,6]+[1,3,4,5,6],
 \end{eqnarray}
 while for the $[2,3\rangle$ shift:
 \begin{eqnarray}
 \frac{A_6^{NMHV}}{A_6^{MHV}}&=&\mathbb{P}\left([1,2,3,4,5]+[1,2,3,5,6]
 +[1,3,4,5,6]\right)\nonumber\\
 &=&[6,1,2,3,4]+[6,1,2,4,5]+[6,2,3,4,5].
\end{eqnarray}
In the general case, the equivalence between different BCFW representations
can be shown using the
representation of the amplitude as an integral over Grassmannian and
residues theorems for functions of multiple complex variables
\cite{Arkani_Hamed_Dual_S_matrix}.
The case $n=6$ may also be seen as the manifestation of six term identity
\begin{eqnarray}
0&=&[1,2,3,4,5]+[1,2,3,5,6]+[1,3,4,5,6]
\nonumber\\&-&\mathbb{P}\left([1,2,3,4,5]+[1,2,3,5,6]
 +[1,3,4,5,6]\right),
 \end{eqnarray}
for the $[a,b,c,d,e]$ functions, which can be interpreted
as " the boundary of 5-simplex in $\mathbb{CP}^4$ =0" in the polytope picture.
In the case of the form factor, we have similar
relations between the $[a,b,c,d,e]$ functions in special kinematics
($\gamma^+=0$). For the $[1,2\rangle$ shift one can get:
\begin{eqnarray}
 \frac{Z_4^{NMHV}}{Z_4^{MHV}}=
 (\mathbb{S}c^{(3)}_2)[-1,0,1,3,4]+[1,2,3,4,5]+[1,2,3,0,-1]+c^{(4)}_2[1,2,3,-2,-1],
 \nonumber\\
 \end{eqnarray}
 while for the $[2,3\rangle$ shift:
\begin{eqnarray}
 \frac{Z_4^{NMHV}}{Z_4^{MHV}}=\mathbb{P}\left((\mathbb{S}c^{(3)}_2)[-1,0,1,3,4]
 +[1,2,3,4,5]
 +[1,2,3,0,-1]+c^{(4)}_2[1,2,3,-2,-1]\right),\nonumber\\
\end{eqnarray}
and as the consequence
\begin{eqnarray}
 0&=&(\mathbb{S}c^{(3)}_2)[-1,0,1,3,4]+[1,2,3,4,5]
 +[1,2,3,0,-1]+c^{(4)}_2[1,2,3,-2,-1]\nonumber\\
 &-&\mathbb{P}\left((\mathbb{S}c^{(3)}_2)[-1,0,1,3,4]
 +[1,2,3,4,5]
 +[1,2,3,0,-1]+c^{(4)}_2[1,2,3,-2,-1]\right).
 \nonumber\\
\end{eqnarray}
Is there any geometrical picture behind such identities
(see also (\ref{3term}))?

It would be interesting to find representations for the form factors
as an integral over Grassmannian \cite{Arkani_Hamed_Dual_S_matrix}
similar to the amplitudes\footnote{Here
$M_i$ is i'th ordered minor of the $n\times k$ $C_{al}$ matrix, and
$\mathcal{W}_l^A=(\mu^{\alpha}_l,\tilde{\lambda}_{\dot{\alpha},l},\eta^A_l)$.} case:
\begin{eqnarray}
 A_n^{(0)(k)}=\int \frac{d^{n\times k}C_{al}}{Vol[GL(k)]}\frac{1}{M_1...M_n}
 \prod_{a=1}^k
 \delta^{4|4}\left(\sum_{l=1}^n C_{al} \mathcal{W}_l^A\right),
\end{eqnarray}
or prove that
such representation is impossible. This representation
is the first step in the on-shell diagram formalism
\cite{Arcani_Hamed_PositiveGrassmannians}, which
may be very useful for the form factors as well as for the amplitudes.
The representation of the ratio of the NMHV
and MHV form factors as the sum of the $[*,a,b,c,d]$ functions gives hope that
such Grassmannian integral representation is possible.

It would be interesting to formulate recursion relations for
the integrand of the form factors at the loop level.
The form factors of operators from
the stress tensor supermultiplet naturally involve non planar contributions
starting from two loops, so to formulate such recursion relations, one
must incorporate non planarity.

And also, it would be interesting to continue the investigation of
the form factors/Wilson loop duality. One can hope that the results obtained
in this
article will be useful in mentioned above quests.

\section*{Acknowledgements}
The author would like to thank D. I. Kazakov,  A. A. Gorsky and
A. V. Zhiboedov for valuable and stimulating discussions. The
author also would like also to thank A.V. Andreyash and S.E. Kuratov
for giving opportunity to finish this
work at the Center for Fundamental and Applied Research.
Financial support RFBR grant \# 14-02-00494 is kindly acknowledged.

\appendix

\section{$\mathcal{N}=4$ harmonic superspaces}
The standard $\mathcal{N}=4$ coordinate superspace
is convenient to describe supermultiplets of fields
or local operators. It is parameterized by the following coordinates:
\begin{equation}
\mbox{$\mathcal{N}=4$ coordinate
superspace}=\{x^{\alpha\dot{\alpha}},~\theta^A_{\alpha},~\bar{\theta}_{A\dot{\alpha}}\},
\end{equation}
where $x_{\alpha\dot{\alpha}}$ are ordinary coordinates,
which are  bosonic variables and
$\theta$'s are additional fermionic coordinates;
$A$ is the $SU(4)_R$ index, $\alpha,\dot{\alpha}$ are the Lorentz $SL(2,C)$ indices.

The $\mathcal{N}=4$ supermultiplet of fields (containing $\phi^{AB}$
scalars, $\psi^A_{\alpha}, \bar{\psi}^A_{\dot{\alpha}}$ fermions and
$F^{\mu\nu}$-- the gauge field strength tensor, all in the adjoint
representation of the $SU(N_c)$ gauge group) is realised in the
$\mathcal{N}=4$ coordinate superspace as the constrained superfield $
W^{AB}(x,\theta,\bar{\theta})$ with the lowest component $
~W^{AB}(x,0,0)=\phi^{AB}(x)$; $W^{AB}$ in general is not a chiral
object and satisfies several constraints: the self-duality constraint
\begin{equation}
W^{AB}(x,\theta,\bar{\theta})
=\overline{W_{AB}}(x,\theta,\bar{\theta})=
\frac{1}{2}\epsilon^{ABCD}W_{CD}(x,\theta,\bar{\theta}),
\end{equation}
which implies $\phi^{AB}=\overline{\phi_{AB}}=\frac 12
\epsilon^{ABCD} \phi_{CD}$ and two additional
constraints\footnote{$[\ast,\star]$ denotes antisymmetrization in
indices, while $(\ast,\star)$ denotes symmetrization in indices.}
\begin{eqnarray}\label{PartialChirality} &&
D_C^{\alpha}W^{AB}(x,\theta,\bar{\theta}) = -\frac{2}{3}\delta^{[A}_CD_L^{\alpha}W^{B]L}(x,\theta,\bar{\theta}), \nonumber\\
&&\bar{D}^{\dot{\alpha}(C}W^{A)B}(x,\theta,\bar{\theta}) = 0,
\end{eqnarray}
where $D^A_{\alpha}$ is the standard coordinate superspace
derivative\footnote{which is
$D^A_{\alpha}=\partial/\partial\theta_{A}^{\alpha}
+i\bar{\theta}^{A\dot{\alpha}}\partial/\partial
x^{\alpha\dot{\alpha}}$.}. Note that in this formulation the full
$\mathcal{N}=4$ supermultiplet of fields is on-shell in the sense
that the algebra (more precisely the last two anticommutators) of the
generators $Q_{A\alpha},\bar{Q}^B_{\dot{\alpha}}$ for the
supersymmetric transformation of the fields in this supermultiplet
\begin{equation}
\{Q_{A\alpha},\bar{Q}^B_{\dot{\alpha}}\}=2\delta^B_AP_{\alpha\dot{\alpha}},~
\{Q_{A\alpha},Q_{B\beta}\}=0,~\{\bar{Q}^A_{\dot{\alpha}},\bar{Q}^B_{\dot{\beta}}\}=0
\end{equation}
is closed only if the fields obey their equations of motion (in
addition the closure of the algebra requires the compensating gauge
transformation \cite{SuperCor1}).

The off-shell formulation of the full $\mathcal{N}=4$ supermultiplet is still
unknown.
But fortunately the self-dual (chiral) sector of the full $\mathcal{N}=4$ supermultiplet
can be formulated off-shell. In the $SU(4)_R$ covariant way this can be done by using the
$\mathcal{N}=4$ harmonic superspace \cite{N=4_Harmonic_SS,SuperCor1}.

The $\mathcal{N}=4$
harmonic superspace is obtained by adding additional bosonic
coordinates (harmonic variables) to the $\mathcal{N}=4$ coordinate
superspace or on-shell momentum superspace. These additional bosonic
coordinates parameterize the coset
\begin{equation}
\frac{SU(4)}{SU(2) \times SU(2)' \times U(1)}
\end{equation}
and carry the  $SU(4)$ index $A$, two copies of the $SU(2)$ indices $a,
\dot{a}$ and the $U(1)$ charge $\pm$
\begin{equation}
(u^{+a}_{A},~u^{-a'}_A)~\mbox{and c.c. once}~(\bar{u}^{-A}_a,~\bar{u}^{+A}_{a'}).
\end{equation}
Using these variables one presents all the Grassmann objects with
$SU(4)_R$ indices. The Grassmann coordinates in the
original $\mathcal{N}=4$ coordinate superspace then can be transformed as
\begin{equation}
\theta^{+a}_{\alpha}=u^{+a}_{A}\theta^A_{\alpha},~~
\theta^{-a'}_{\alpha}=u^{-a'}_{A}\theta^A_{\alpha},
\end{equation}
\begin{equation}
\bar{\theta}^{-}_{a\dot{\alpha}}=\bar{u}^{-A}_a\bar{\theta}_{A\dot{\alpha}},~~
\bar{\theta}^{+}_{a'\dot{\alpha}}=\bar{u}^{+A}_{a'}\bar{\theta}_{A\dot{\alpha}},
\end{equation}
and in the opposite direction
\begin{equation}
\theta^A_{\alpha}=\theta^{+a}_{\alpha}\bar{u}^{-A}_a+
\theta^{-a'}_{\alpha}\bar{u}^{+A}_{a'},
\end{equation}
\begin{equation}
\bar{\theta}_{A\dot{\alpha}}=\bar{\theta}^{+}_{a'\dot{\alpha}}u^{-a'}_A+
\bar{\theta}^{-}_{a\dot{\alpha}}u^{+a}_{A}.
\end{equation}
The same is true for supercharges:
\begin{equation}
Q_{A\alpha}\rightarrow(Q^-_{a\alpha},~Q^+_{a'\alpha}),
~\bar{Q}^{A}_{\dot{\alpha}}\rightarrow(\bar{Q}^{+a}_{\dot{\alpha}},~\bar{Q}^{-a'}_{\dot{\alpha}}).
\end{equation}
So the $\mathcal{N}=4$ harmonic superspace is parameterized with the
following set of coordinates
\begin{eqnarray}
\mbox{$\mathcal{N}=4$ harmonic
superspace}&=&\{x^{\alpha\dot{\alpha}},
~\theta^{+a}_{\alpha},
~\theta^{-a'}_{\alpha},
~\bar{\theta}^{-}_{a\dot{\alpha}},
~\bar{\theta}^{+}_{a'\dot{\alpha}}
~u
\}.\nonumber\\
\end{eqnarray}
Using $u$ harmonic variables one can project the $W^{AB}$ superfield
as
\begin{eqnarray}
W^{AB}\rightarrow W^{AB}u^{+a}_{A}u^{+b}_{B}=\epsilon^{ab}W^{++},
\end{eqnarray}
\begin{eqnarray}
W^{++}=W^{++}(x,~\theta^{+a},
\theta^{-a'},
~\bar{\theta}^{-}_{a},
\bar{\theta}^{+}_{a'}
~u),
\end{eqnarray}
where $\epsilon^{ab}$ is an $SU(2)$ totally antisymmetric tensor.
This $W^{++}$ superfield is $SU(4)_R$ and $SU(2) \times SU(2)' \times U(1)$
covariant but carries $+2$ $U(1)$ charge.

Using harmonics one can project constraints (\ref{PartialChirality})
so that\footnote{Strictly speaking, this is true only in the free theory
($g=0$), in the interacting theory one has to replace
$D_{\alpha}^A,\bar{D}_{\dot{\alpha}}^A$ by their gauge covariant
analogs, which contain superconnection, but the final result is the
same \cite{SuperCor1}.}:
\begin{eqnarray}
D^{\alpha}_{-a'}W^{++}&=&0,\nonumber\\
\bar{D}^{\dot{\alpha}}_{+a}W^{++}&=&0.
\end{eqnarray}
Thus, the superfield $W^{++}$ contains the dependence on half of the
Grassmannian variables $\theta$'s and $\bar{\theta}$'s:
\begin{eqnarray}
W^{++}=W^{++}(x, ~\theta^{+a},
\bar{\theta}_{a'}^{-},~u).
\end{eqnarray}
Now one can put all $\bar{\theta}=0$ in $W^{++}$, the corresponding
supercharges ect. and observe that all component fields in $W^{++}(x, ~\theta^{+a},
0,~u)$ are off-shell in a sense that the remaining chiral part of SUSY algebra
$\{Q_{A\alpha},Q_{B\beta}\}=0$ which acts on $W^{++}$ is closed without using equation of motion for the component fields.

The chiral part $\mathcal{T}$ of the stress tensor supermultiplet
can now be constructed simply as:
\begin{eqnarray}
\mathcal{T}(x,\theta^+,u)=Tr(W^{++}W^{++})|_{\bar{\theta}=0}.
\end{eqnarray}
$\mathcal{T}$ is the first operator in the series of the so-called
1/2-BPS operators of the form $Tr[(W^{++})^k]$.
Its lowest component is
\begin{eqnarray}
\mathcal{T}(x,0,u)=Tr(\phi^{++}\phi^{++}),
~\phi^{++}=\frac{1}{2}\epsilon_{ab}u^{+a}_{A}u^{+b}_{B}\phi^{AB},
\end{eqnarray}
and its highest component which is proportional to $(\theta^+)^4$ is
the Lagrangian of $\mathcal{N}=4$ SYM written in a special (chiral) form.
All components of
$\mathcal{T}$ can be found in \cite{SuperCor1}. Using supercharges one
can write $\mathcal{T}$ as:
\begin{eqnarray}
\mathcal{T}(x,\theta^+,u)=exp(\theta^{+a}_{\alpha}Q^{-\alpha}_{a})Tr(\phi^{++}\phi^{++}).
\end{eqnarray}
Also, the lowest component $\mathcal{T}(x,0,u)$ commutes with
half of the chiral and anti-chiral supercharges of the theory:
\begin{eqnarray}
[\mathcal{T}(x,0,u),Q^{+}_{a'\alpha}]=0,
~[\mathcal{T}(x,0,u),\bar{Q}^{+a}_{\dot{\alpha}}]=0.
\end{eqnarray}
These properties allow one to
determine the general Grassmann structure of the form factor
\cite{BKV_SuperForm}.

Harmonic variables can also be used in on-shell momentum superspace to treat
on-shell states of the theory on equal footing as operators from supermultiplets.
Using harmonic variables one can write:
\begin{eqnarray}
\mbox{$\mathcal{N}=4$ harmonic
on-shell momentum superspace}&=&\{\lambda_{\alpha},\tilde{\lambda}_{\dot{\alpha}},~\eta^{-}_a,\eta^{+}_{a'},~u\}.
\nonumber\\
\end{eqnarray}
Here $\lambda_{\alpha}$ and $\tilde{\lambda}_{\dot{\alpha}}$ are the $SL(2,C)$ spinors associated with momenta carried by a massless state (particle): $p_{\alpha\dot{\alpha}}=\lambda_{\alpha}\tilde{\lambda}_{\dot{\alpha}}$, $p^2=0$.
Supercharges which act in this superspace can be
represented in the n-particle case as
\begin{eqnarray}
q_{a\alpha}^-=\sum_{i=1}^n\lambda_{\alpha,i}\eta^-_{a,i},
~~q_{a'\alpha}^+=\sum_{i=1}^n\lambda_{\alpha,i}\eta^+_{a',i},
\end{eqnarray}
and
\begin{eqnarray}
\bar{q}_{\dot{\alpha}}^{+a}=\sum_{i=1}^n\tilde{\lambda}_{\dot{\alpha},i}\frac{\partial}{\eta^-_{a,i}},
~~\bar{q}_{\dot{\alpha}}^{-a'}=\sum_{i=1}^n\tilde{\lambda}_{\dot{\alpha},i}\frac{\partial}{\eta^+_{a',i}}.
\end{eqnarray}

The Grassmann delta functions, which one can encounter in this article, are given by
($ \langle ij \rangle\equiv\lambda_{\alpha,i}\lambda_{j}^{\alpha}$):
\begin{eqnarray}
\delta^{-4}(q_{a\alpha}^-)=\sum_{i,j=1}^n\prod_{a,b=1}^2\langle ij \rangle \eta^{-}_{a,i}\eta^{-}_{b,j},~~
\delta^{+4}(q_{a\alpha}^+)=\sum_{i,j=1}^n\prod_{a',b'=1}^2\langle ij \rangle
\eta^{+}_{a',i}\eta^{+}_{b',j},
\end{eqnarray}
\begin{eqnarray}
\hat{\delta}^{-2}(X^{-a})=\prod_{a=1}^2X^{-a},~~\hat{\delta}^{+2}(X^{+}_{a'})=\prod_{a=1}^2X^{+}_{a'}.
\end{eqnarray}
We also will use the notations
\begin{eqnarray}
\delta^{-4}\delta^{+4}\equiv\delta^{8},~\hat{\delta}^{-2}\hat{\delta}^{+2}\equiv\hat{\delta}^4.
\end{eqnarray}
Using these delta functions one can rewrite the $\mbox{MHV}_n$ and $\overline{\mbox{MHV}}_3$ amplitudes, $R_{rst}$
functions etc. in the form nearly identical to
the form they have in the ordinary on-shell momentum superspace.

Grassmann integration measures are defined as
\begin{eqnarray}
d^{-2}\eta=\prod_{a=1}^2d\eta_a^{-},
~d^{+2}\eta=\prod_{a=1}^2d\eta^{+a'},
~d^{-2}\eta d^{+2}\eta\equiv d^4\eta.
\end{eqnarray}
in the on-shell momentum superspace and
\begin{eqnarray}
d^{-4}\theta=\prod_{a,\alpha=1}^2d\theta^{-a}_{\alpha},
~d^{+4}\theta=\prod_{a,\alpha=1}^2d\theta^{+}_{a'\alpha},
\end{eqnarray}
in the ordinary superspace; $\delta^{\pm4}$ functions can be represented as $\hat{\delta}^{\pm2}$
functions using the identity (here we drop the $SU(2)$ and $SL(2,C)$ indices),
\begin{equation}\label{supersumm_delta_to_hatt_deltas}
\delta^{\pm4}(q^{\pm})=\langle lm
\rangle^{2}\hat{\delta}^{\pm2}\left(\eta^{\pm}_{l}+\sum_{i=1}^n\frac{\langle
mi \rangle}{\langle ml
\rangle}\eta_i^{\pm}\right)\hat{\delta}^{\pm2}\left(\eta^{\pm}_{m}+\sum_{i=1}^n\frac{\langle
li \rangle}{\langle lm \rangle}\eta_i^{\pm}\right), ~i \neq l,~i\neq m.
\end{equation}
which can be integrated as usual Grassmann delta functions.

\section{Spurious pole cancellation in $A_5^{NMHV(0)}$ and $Z_4^{NMHV(0)}$}
Now let us illustrate how the cancellation of the spurious poles can be seen
on the example of the $\mbox{NMHV}_5$ amplitude.
Consider the all-line shift (CSW) representation of the
$\mbox{NMHV}_5$ amplitude:
\begin{eqnarray}
\frac{A^{NMHV}_5}{A^{MHV}_5}=[*,1,2,3,4]+
[*,2,3,4,5]+[*,3,4,5,1]+[*,4,5,1,2]+[*,5,1,2,3].\nonumber\\
\end{eqnarray}
Applying the boundary operator to all terms in $A^{NMHV}_5/A^{MHV}_5$ we get:
\begin{eqnarray}
\partial Vol_4[*,1,2,3,4]&=&\nonumber
Vol_3[1,2,3,4]-Vol_3[*,2,3,4]+Vol_3[*,1,3,4]-Vol_3[*,1,2,4]
\nonumber\\&+&Vol_3[*,1,2,3],\nonumber\\
\partial Vol_4[*,2,3,4,5]&=&\nonumber
Vol_3[2,3,4,5]-Vol_3[*,3,4,5]+Vol_3[*,2,4,5]-Vol_3[*,2,3,5]
\nonumber\\&+&Vol_3[*,2,3,4],\nonumber\\
\partial Vol_4[*,3,4,5,1]&=&\nonumber
Vol_3[3,4,5,1]-Vol_3[*,1,4,5]+Vol_3[*,1,3,5]-Vol_3[*,1,3,4]
\nonumber\\&+&Vol_3[*,3,4,5],\nonumber\\
\partial Vol_4[*,4,5,1,2]&=&\nonumber
Vol_3[4,5,1,2]-Vol_3[*,1,2,5]+Vol_3[*,1,2,4]-Vol_3[*,2,4,5]
\nonumber\\&+&Vol_3[*,1,4,5],\nonumber\\
\partial Vol_4[*,5,1,2,3]&=&\nonumber
Vol_3[5,1,2,3]-Vol_3[*,1,2,3]+Vol_3[*,2,3,5]-Vol_3[*,1,3,5]
\nonumber\\&+&Vol_3[*,1,2,5].\nonumber\\
\end{eqnarray}
We see that all terms containing $\mathcal{Z}_{*}$ "cancel" each other, which
indicates that in the sum of all terms all
spurious poles $\langle *,a,b,c \rangle$
are canceled.

Now let us consider the $\mbox{NMHV}_4$ form factor. In the
all-line shift (CSW) representation it can be written as:
\begin{eqnarray}
Z^{NMHV}_4/Z^{MHV}_4&=&([*,-1,0,1,2]+[*,-1,0,2,3])+([*,0,1,2,3]+[*,0,1,3,4])+
\nonumber\\
&+&([*,1,2,3,4]+[*,1,2,4,5])+([*,2,3,4,5]+[*,2,3,5,6]).\nonumber\\
\end{eqnarray}
Note also that equivalently one can rewrite last two terms as
\begin{eqnarray}
[*,2,3,4,5]=[*,-2,-1,0,1],~[*,2,3,5,6]=[*,-2,-1,1,2].
\end{eqnarray}
Applying $\partial$ to all these terms one can obtain:
\begin{eqnarray}
\partial Vol_4[*,-1,0,1,2]&=&Vol_3[-1,0,1,2]-Vol_3[0,1,2,*]+Vol_3[1,2,*,-1]-Vol_3[2,*,-1,0]
\nonumber\\&+&Vol_3[*,-1,0,1],\nonumber\\
\partial Vol_4[*,-1,0,2,3]&=& Vol_3[-1,0,2,3]-Vol_3[0,2,3,*]+(Vol_3[2,3,*,-1]-Vol_3[3,*,-1,0])
\nonumber\\&+&Vol_3[*,-1,0,2],\nonumber\\
\partial Vol_4[*,0,1,2,3]&=&Vol_3[0,1,2,3]-Vol_3[1,2,3,*]+Vol_3[2,3,*,0]-Vol_3[3,*,0,1]
\nonumber\\&+&Vol_3[*,0,1,2],\nonumber\\
\partial Vol_4[*,0,1,3,4]&=& Vol_3[0,1,3,4]-Vol_3[1,3,4,*]+(Vol_3[3,4,*,0]-Vol_3[4,*,0,1])
\nonumber\\&+&Vol_3[*,0,1,3],\nonumber\\
\partial Vol_4[*,1,2,3,4]&=&Vol_3[1,2,3,4]-Vol_3[2,3,4,*]+Vol_3[3,4,*,1]-Vol_3[4,*,1,2]
\nonumber\\&+&Vol_3[*,1,2,3],\nonumber\\
\partial Vol_4[*,1,2,4,5]&=& Vol_3[1,2,4,5]-Vol_3[2,4,5,*]+(Vol_3[4,5,*,1]-Vol_3[5,*,1,2])
\nonumber\\&+&Vol_3[*1,2,4],\nonumber\\
\partial Vol_4[*,-2,-1,0,1]&=&Vol_3[-2,-1,0,1]-Vol_3[-1,0,1*]+Vol_3[0,1,*,-2]-Vol_3[1,*,-2,-1]
\nonumber\\&+&Vol_3[*,-2,-1,0],\nonumber\\
\partial Vol_4[*,-2,-1,1,2]&=& Vol_3[-2,-1,1,2]-Vol_3[-1,1,2,*]+(Vol_3[1,2,*,-2]-Vol_3[2,*,-2,-1])
\nonumber\\&+&Vol_3[*,-2,-1,1].\nonumber\\
\end{eqnarray}
We see that poles corresponding to terms containing
$\mathcal{Z}^*$ in the $(\ldots)$
bracket "cancel" in $\gamma^+ \to 0$ limit, while all other
$\mathcal{Z}^*$ dependant poles "cancel" among
themselves.

\section{IR pole coefficients relations}
In one loop generalized unitarity based calculations for the $\mbox{NMNV}$
sector the following
identities for the $R$ functions were used in $n=4$ case:
\begin{eqnarray}
\tilde{R}_{244}^{(1)}=\tilde{R}_{211}^{(1)},
~\tilde{R}_{144}^{(1)}=\tilde{R}_{311}^{(1)},
~R^{(2)}_{413}=R^{(1)}_{241}.
\end{eqnarray}
We now want to show that they are transparent and easily derived
in the momentum twistor variables.

Let us start with
$
\tilde{R}_{244}^{(1)}=\tilde{R}_{211}^{(1)}.
$
It is essentially trivial, these are the same $R$
functions written using clockwise and anticlockwise conventions.

For
$
\tilde{R}_{144}^{(1)},~\tilde{R}_{311}^{(1)},
$
one can obtain (note that here legs are ordered clockwise )
\begin{eqnarray}
\tilde{R}_{144}^{(1)}=\frac{\langle 1,2,4,-1 \rangle\langle 4,-1,0,1 \rangle}
{\langle -1,0,3,4 \rangle\langle 1,3,4,5 \rangle}
[1,3,4,-1,0]=[*,0,1,3,4]+[*,-1,0,1,3]+[*,1,3,4,5].\nonumber\\
\end{eqnarray}
\begin{eqnarray}
\tilde{R}_{311}^{(1)}=\frac{\langle 3,4,5,0 \rangle\langle 5,0,1,3 \rangle}
{\langle 0,1,4,5 \rangle\langle 3,-1,0,1 \rangle}
[3,4,5,0,1]=[*,0,1,3,4]+[*,1,3,4,5]+[*,3,-1,0,1].\nonumber\\
\end{eqnarray}
Indeed, as expected $\tilde{R}_{144}=\tilde{R}_{311}$.

For
$R^{(2)}_{413}$ and $R^{(1)}_{241}$,
we see that
\begin{eqnarray}
R^{(2)}_{413}=[4,0,1,2,3],
~\mbox{and}~R^{(1)}_{241}=[2,3,4,0,1],
\end{eqnarray}
so $R^{(2)}_{413}=R^{(1)}_{241}$ as expected.

\end{document}